\documentclass[11pt]{article}
\usepackage{amsmath,amssymb,graphicx,color,epsfig,setspace}
\usepackage[square, comma,sort&compress]{natbib}

\begin{document}

\doublespacing

\title{Accurate path integration in continuous attractor network models of grid cells
\footnote{Burak Y, Fiete IR, 2009. PLoS Comput Biol 5(2): e1000291. doi:10.1371/journal.pcbi.1000291}}
\author{Yoram Burak\thanks{E-mail:yburak@fas.harvard.edu}$^{\dagger,1,2}$ 
and Ila R. Fiete\thanks{E-mail: ilafiete@caltech.edu}$^{\ddagger,2,3}$\\
$^{1}${Center for Brain Science, Harvard University, Cambridge, MA 02138}\\
$^{2}${Kavli Institute for Theoretical Physics, UCSB, Santa Barbara, CA 93106, USA}\\
$^{3}${Computation and Neural Systems, Division of Biology, Caltech, Pasadena CA 91125}}
\date{January 3, 2009}

\maketitle
\begin{abstract}
Grid cells in the rat entorhinal cortex display strikingly regular firing
responses to the animal's position in 2-D space, and have been hypothesized to form the neural substrate for dead-reckoning. However, errors
accumulate rapidly when velocity inputs are integrated in existing models of grid cell activity. To produce grid-cell like responses, these models would require frequent resets triggered by external sensory cues. Such inadequacies, shared by various models, cast doubt
on the dead-reckoning potential of the grid cell system. Here we focus on the question of accurate path integration, specifically in continuous attractor models of grid cell activity. We show, in contrast to previous models, that continuous attractor models can generate regular triangular grid responses, based on
inputs that encode only the rat's velocity and heading direction. We consider the role of the network boundary in the integration performance of the network, and show that both periodic and aperiodic networks are capable of accurate path integration, despite important differences in their attractor manifolds. We quantify the rate at which errors in the velocity integration accumulate as a function of network size and intrinsic noise within the
network. With a plausible range of parameters and the inclusion of spike variability, our model networks can accurately integrate velocity inputs over a maximum of $\sim 10 - 100$ m and $\sim 1-10$ minutes. 
These findings form a proof-of-concept that continuous attractor dynamics may underlie velocity integration in dMEC. 
The simulations also generate pertinent upper bounds on the accuracy of integration that may be achieved by continuous attractor dynamics in the grid cell network. 
We suggest experiments to test the continuous attractor model and differentiate it from models in which single cells establish their responses independently of each other.
\end{abstract}
\section*{Author Summary}
Even in the absence of external sensory cues, foraging rodents maintain an estimate
of their position, allowing them to return home in a roughly
straight line. This computation is known as dead reckoning or path integration. A discovery made three years ago in rats focused
attention on the dorsolateral medial entorhinal cortex (dMEC) as a
location in the rat's brain where this computation might be performed.
In this area, so-called grid cells fire whenever the rat is on any
vertex of a triangular grid that tiles the plane. Here we propose a
model that could generate grid-cell-like responses in a neural
network. The inputs to the model network convey information about the rat's
velocity and heading, consistent with known inputs projecting into
dMEC. The network effectively integrates these inputs to produce a
response that depends on the rat's absolute position. We show that
such a neural network can integrate position accurately, and can
reproduce grid-cell like responses similar to those observed
experimentally. We then suggest a set of experiments that could help
identify whether our suggested mechanism is responsible for the
emergence or grid cells, and for path integration in the rat's brain.
\section*{Introduction}
Since the discovery of grid cells in the dorsolateral band of the medial entorhinal cortex (dMEC) \citep{Hafting05}, several ideas have been put forth on how grid-cell activity might emerge  \citep{Fuhs06,Burak06,McNaughton06,Burgess07,Hasselmo07,Guanella07}. The theoretical ideas suggested so far fall into two categories. In continuous attractor models (see \citep{Skaggs95,Seung96,Zhang96,Samsonovich97,Tsodyks99,Goodridge00,Xie02,Stringer04} and \citep{Fuhs06, McNaughton06, Guanella07} for the grid cell system), 
which are the focus of this work, grid cell activity arises from the collective behavior of a neural network. The network's state is restricted to lie in a low-dimensional continuous manifold of steady states, and its particular location within this manifold is updated in response to the rat's velocity. In the second category of models \citep{OKeefe05,Burgess07,Hasselmo07,Giocomo07}, grid-cell activity arises independently in single cells, as a result of interference between a global periodic signal and a cell-specific oscillation, whose frequency is modulated by the rat's velocity.

These ideas differ radically from each other, but they share a common assumption about the nature of the input feeding into dMEC, namely, that the input conveys information primarily on the rat's velocity and heading. Within all these models, grid cell activity must then arise from precise integration of the rat's velocity.

Grid cell firing exhibits remarkable accuracy: The periodic spatial tuning pattern remains sharp and stable over trajectories lasting 10's of minutes, with an accumulated length on the order of hundreds of meters \citep{Hafting05}. Experiments performed in the dark show that grid cell tuning remains relatively accurate over $\sim$100 meters and $\sim$10 minutes even after a substantial reduction of external sensory inputs. However, in these experiments olfactory and tactile cues were not eliminated, and grid cell responses may have been informed by positional information from such cues. Therefore, the duration and length of paths over which coherent grid responses are maintained without any external sensory cues is not known. For position estimation on the behavioral level, we searched for but found no clear quantitative records of the full range over which rats are capable of accurate dead-reckoning. Behavioral studies \citep{Mittelstaedt80, Maaswinkel99,Sharp01,Etienne04} document that rats can compute the straight path home following random foraging trajectories that are 1-3 meters in length, in the absence of external sensory cues.

How do theoretical models measure up, in estimating position from input velocity cues? The theta-oscillation model of grid cells \citep{OKeefe05,Burgess07,Hasselmo07,Giocomo07}, under idealized assumptions about internal connectivity, velocity inputs, and neural dynamics, is not able to produce accurate spatial grids over the known length- and time-scales of behavioral dead-reckoning if the participating theta oscillations deviate from pure sine waves. This is because the model is acutely vulnerable to subtle changes in the phase of the underlying oscillations. In reality, theta oscillations are not temporally coherent: cross-correlograms from {\em in vitro} intracellular recordings \citep{Alonso89,Alonso93,Giocomo07} and {\em in vivo} extracellular recordings \citep{Mitchell80,Alonso87} show that the phase of the theta oscillation in the entorhinal cortex typically decoheres or slips by half a cycle in less than 10 cycles or about 1 second, which corresponds to a distance of only 1 meter for a run velocity of 1 m/s. This means that the model grid cells will entirely lose track of the correct phase for the present rat position within that time. 

For continuous attractor models, we previously showed \citep{Burak06} that due to rotations and non-linear, anisotropic velocity responses, a detailed model \citep{Fuhs06} integrates velocity poorly, and does not produce a grid-cell firing pattern even with idealized connectivity and deterministic dynamics. 
Another model \citep{Guanella07} generates grid responses in a small periodic network, but it includes no neural nonlinearities or variability in neural responses, and depends on real-time, continuous modulation of recurrent weights by the velocity inputs to the network.

Conceptually, the existence of an integrating apparatus seems pointless if it is completely dependent on nearly continuous corrections coming from an external source that specifies absolute position. Thus, it seems reasonable to require that theoretical models of path integration in dMEC, if using faithful velocity inputs, have the ability to reproduce stable grid cell patterns for trajectories lasting a few minutes. 

Our aim, therefore, is to establish whether it is possible for model grid cells to accurately integrate velocity inputs. We restrict our analysis specifically to continuous attractor networks. As will become clear, the precision of velocity integration can strongly depend on various factors including network topology, network size, variability of neural firing, and variability in neural weights. Here we focus on three of these factors: boundary conditions in the wiring of the network (periodic vs. aperiodic), network size, and stochasticity in neural activity 

We quantify path integration accuracy in both periodic and aperiodic recurrent network models of dMEC, and demonstrate that within a biologically plausible range of parameters explored, such networks have maximum attainable ranges of accurate path integration of 1-10 minutes and 10-100 meters. Larger, less noisy networks occupy the high end of the range, while smaller and more stochastic networks occupy the low end.
We end with suggestions for experiments to quantify integration accuracy, falsify the continuous attractor hypothesis, and determine whether the grid cell response is a recurrent network phenomenon or whether it emerges from computations occurring within single cells. 

\section*{Results}

In our model, each neuron receives inhibitory input from a surrounding ring of local neurons. The entire network receives broad-field feedforward excitation (\textit{Methods}). If the inhibitory interactions are sufficiently strong, this type of connectivity generically produces a population response consisting of a regular pattern of discrete blobs of neural activity, arranged on the vertices of a regular triangular lattice \citep{Murray_book, Burak06,McNaughton06}, Figure 1A. Ignoring boundary effects for the moment, all possible phases (translations) of the pattern are equivalent steady states of the pattern formation process, and therefore form a continuous attractor manifold.

To reproduce the regular single-neuron (SN) lattice patterns observed in experiment, the pattern formed in the neural population must be coupled to the rat's velocity. This coupling is arranged in such a way (Figure 1B and \textit{Methods}) that it drives translations of the pattern within the neural sheet, in proportion to the movements of the rat in real 2-d space, Figure 1C. 

Briefly, velocity coupling involves distributing a set of direction labels ($\theta$) to the neurons in any patch of the network (Figure 1B). The direction label $\theta$ signifies that (1) the neuron receives input from a speed-modulated head-direction cell tuned to that direction, and (2) the neuron's outgoing center-surround connectivity profile is centered not on itself, but is shifted by a few neurons along a corresponding direction on the neural sheet. The neuron tends, through its slightly asymmetric connectivity, to drive network activity in the direction of the shift. However, another neuron with the opposite direction preference will tend to drive a flow in the opposite direction. If all neurons have equal inputs, the opposing drives will balance each other, and the activity pattern will remain static. If, however, the rat moves in a particular direction in space, the corresponding model dMEC cells will receive larger input than the others, due to their head-direction inputs, and will succeed in driving a flow of the network pattern along their preferred direction. 
This mechanism for input-driven pattern flow is similar to that proposed in a model of the head-direction system \citep{Xie02}.
Figure 1C demonstrates how a flow of the population pattern will drive activity 
at spatially periodic intervals in single neurons. 

To obtain spatially periodic responses in single neurons over long, curved, variable-speed trajectories, additional conditions must be met, as we discuss below. We present results from two topologically distinct networks: one with aperiodic, and the other with periodic, connectivity.

\subsection*{A periodic network accurately integrates rat velocity}
We simulate dynamics in a network of neurons driven by velocity inputs obtained from recordings of a rat's trajectory (see \textit{Methods}). The network contains $128^2$ ($\sim 10^4$) neurons arranged in a square sheet. Neurons close to each edge of the sheet form connections with neurons on the opposite edge, such that the topology of the network is that of a torus. Figure 2A shows the population activity in the network at one instant of the run. 

A grid cell response, as reported in experimental papers, is obtained by summing the firing activity of a single neuron over a full trajectory. Unlike the population response, which is an instantaneous snapshot of full neural population, the single-neuron response is an integrated measure over time of the activity one cell. In the rest of this paper, SN response refers to the accumulated response of single neurons over a trajectory.

In the periodic network, the SN response, accumulated over the $\sim 20$ minute  trajectory, and plotted as a function of the true rat position, shows coherent grid activity, Figure 2\,B. The network accurately integrates input velocity, as can verified directly by comparing the cumulative network pattern phase to the rat's true position, Figure 2C. The total error, accumulated over $\sim 260$ m and 20 minutes, is $< 15$ cm, compared to a grid period of about 48 cm. This corresponds to an average integration error of less than 0.1 cm per meter traveled and less than 0.01 cm per second traveled. The range of rat speeds represented in the input trajectory was 0-1 m/s, showing that this network is capable of accurate path integration over this range of speeds.  

A deterministic periodic network of only $40^2$ ($\sim 10^3$) neurons also performs well enough to produce coherent SN grids over the same trajectory, Figure S1. 

\subsection*{Equivalent conditions for accurate path integration}

The presence of a clear spatial grid in the SN response to velocity inputs alone is a good indication of the accuracy of integration. If the rat's internal estimate of position were to drift by half a grid period, the neuron would fire in the middle of two existing vertices rather than on a vertex. As the rat traveled over its trajectory, the neuron would fire at various ``wrong'' locations, with the resulting SN response becoming progressively blurred until no grid would be discernible. This would happen even if the population pattern remained perfectly periodic throughout. 

Therefore, the following properties are equivalent: (1) Coherent grids in the SN responses, (2) Accurate path integration of the full trajectory over which the SN responses are visualized, with errors smaller than the grid period. An example of this equivalence is given in Figures 2A, C, which show sharp SN patterning and a very small integration error. 

Next, because the population pattern phase accumulates errors whenever the pattern slips relative to rat motion, another equivalent condition for accurate path integration is (3) Linear relationship between network flow velocity and input velocity over the input velocity range, independent of direction.

These equivalent conditions for accurate integration apply to both periodic and aperiodic network models of grid cells (discussed next).

\subsection*{An appropriately configured aperiodic network can accurately integrate rat velocity} 
It is unclear whether a torus-like network topology, in which neurons along opposite edges of the network are connected to form periodic boundary conditions, exists in the rat's brain. Even if such connectivity exists, it may require, at an earlier stage of development, an initially aperiodic network (see Discussion).
Hence it is interesting to consider whether a network with non-periodic boundaries can produce grid-cell like SN activity. The difficulty here is that as the population pattern flows in response to velocity inputs, it must reform
at the boundaries of the neural sheet. Newly forming activity blobs must be created at accurate positions, and the process must not interfere with the pattern's flow.

A central result of the present work on aperiodic networks is that such networks can, in fact, accurately integrate velocity inputs. With an appropriate choice of architecture and inputs and with deterministic dynamics, an aperiodic network can produce SN responses that are as accurate as in the periodic case above. This is illustrated in the example of Figure 2\,D--F. At the aperiodic boundaries, the same dynamics that governed the initial pattern formation process also cause the pattern to continually regenerate as the pattern flows (Figure 1C, bottom). The phases or locations of the renewing blobs at the boundary are consistent with the rest of the network pattern, in part because their placement is influenced by inhibition from the neighboring active neurons in the network interior.

\subsection*{Accurate integration in aperiodic networks is not generic.} 

Despite the success of the model given above, 
accurate path integration in aperiodic networks is not as generic an outcome as it was in the periodic network. We describe next how accurate path integration in aperiodic networks requires attention to details and tuning. 

To produce a coherent SN grid in aperiodic networks, as above, it is not sufficient to simply leave unconnected 
the opposite edges of the sheet that were connected together to produce a periodic network: If the recurrent connections and external inputs terminate abruptly at the network edge, 
the population activity pattern there becomes severely distorted. Such distortions disrupt the linearity of the network's response to velocity inputs \citep{Burak06}. As a result, population pattern distortions, even when confined to the edges of the network, globally destroy the possibility of generating grid-like SN responses for any neuron, including those in the interior of the network where the pattern is locally undistorted.  In fact, even subtle distortions of the pattern near the edges cause similar problems. 

{\bf Modulation of recurrent weights vs. feedforward inputs.} To ameliorate the problem of edge distortions, we considered two main types of modulation in the network architecture. One of these, as in \citep{Fuhs06}, was to smoothly modulate the strength of weights to zero near the boundary. Generally speaking, this method still leads to distorted patterning near the edges. To see why, consider that if weights are sufficiently weak, then pattern formation, which is driven by the recurrent connectivity, does not occur at all. The uniform mode, in which all neurons are equally active, becomes stable. Thus fading the strength of recurrent connectivity to small values at the boundaries results in distortions of the triangular lattice pattern, including the formation of a band of uniformly and highly active neurons along the edges  (\citep{Fuhs06,Burak06} and Figure S2). Other modulations of the weights at the edges create other types of mismatch between the pattern at the edges compared to the interior. 

A second approach is to keep the strength of local recurrent connectivity, which is responsible for pattern formation, constant throughout the network and at the edges, while tapering the strength of external feedforward inputs near the edges. The result is that local patterning is robust, but at the same time, neurons in boundary blobs are proportionally less active, with their activation profiles fading smoothly to zero near the network edges. It is straightforward to see, analytically, that if the network dynamics of Eq. 1 has a particular spatially patterned solution $\bar{s}$ (designating the population activity vector) for a given strength of input $B$, the solution for the scaled input vector $\gamma B$ is the same spatial pattern, scaled in amplitude to $\gamma {\bar s}$. Thus, if the weakening of external inputs is sufficiently gradual (compared to the spacing between activity blobs in the population pattern), activity must scale in proportion to the external input, without a disruption in the periodicity of the pattern. Because the activity of blobs at the network boundary is far lower than in the interior, these boundary blobs have correspondingly less influence on overall network dynamics during flow, and have a less disruptive effect on the linearity of the network response to velocity inputs.

Indeed, we found in our simulations that tapered input profiles dramatically improve the linearity of response to velocity inputs, compared to a modulation of the weights. Throughout the manuscript, therefore, we have used a tapered input profile with untapered weights. An example of faithful population patterning with tapered input can be seen in Figure 2\,D, with the input profile plotted above the population activity.

As we describe next, a population response that appears regular near the edges is necessary, but not sufficient, for accurate integration. 

{\bf Independent effects of network size and input profile on integration accuracy.} 
The input envelope of Figure 3 B is somewhat sharper than in Figure 2, yet is still smooth enough to produce a regular population pattern without irregularities, and with boundary neurons that are only weakly active (Figure 3B). However, this network fails to produce a periodic structure in the SN response (Figure 3A). 
Recording the population activity at different times reveals that the 
population pattern rotates (Figure 3B, C). The velocity inputs, 
which are supposed to drive only pure translation of the pattern, 
also induce rotation. 
Another reason for the network's poor performance is demonstrated in
Figure 3 D2: The flow rate of the 
grid pattern is not precisely proportional to the rat's velocity. 
In particular, at rat velocities below approximately 10 cm/s 
there is no flow at all, and the pattern is ``pinned''. 

The network's ability to produce accurate path integration and coherent SN grids is independently influenced by two factors, the activity profile of neurons at the network boundary, and network size. For a fixed network size, sharper input fading at the boundaries leads to more pinning (Figure 3, D3 \textit{vs.} D2 \textit{vs.} D1). Thus, a relatively subtle difference in how activity fades near the network boundary is sufficient to cause a transition from accurate path integration and coherent SN grids into poor tracking and the complete absence of SN grids. 
At the same time, for a given tapering of inputs at the boundary, increasing the size of the network reduces pinning and improves the linearity of the network velocity response (Figure 3, D4 \textit{vs.} D2), suggesting that from the point of view of integration performance, the larger the network the better. 

The same factors that reduce pinning (smoother input fading at network boundaries and larger network size) also serve to stabilize the orientation of the population pattern (data not shown), suggesting that the undesirable coupling of velocity inputs to rotation is also related to the existence of the boundaries. 
 
A network with $128^2$ ($\sim 10^4$) neurons (Figures 2D-F and Figure 3\,D1,D4) can be large enough, with deterministic dynamics and appropriately chosen boundaries, to perform accurate path integration over 260 m and 20 minutes. Although we did not strenuously attempt to optimize all parameters involved, within our explorations we were unable to construct an aperiodic network substantially smaller than $10^4$ neurons which performs comparably well. It appears, therefore, that network size strongly constrains the accuracy of integration in aperiodic networks, to a greater extent than in the periodic case. 

\subsection*{The attractor manifold}

For the two types of networks from the previous section, the structure of the state-space
is schematically illustrated in Figure 4. The state-space illustration is instrumental in synthesizing the findings of the preceding section -- in particular:  Why does the pattern not rotate in the periodic network? Why is the pattern pinned at low input velocities in the aperiodic network? Why does network size matter more for aperiodic than for periodic networks?  We assume that the dynamics minimize an energy functional, 
whose local minima correspond a set of fixed points (attractors) (This assumption is precisely correct in the absence of a velocity-driven shift mechanism, since the connectivity matrix is then symmetric \citep{Cohen83,Hopfield84}.)

Consider first the \textit{periodic network}. Starting from a steady state of the dynamics, and rigidly translating the stable population pattern, produces an equivalent steady state with exactly the same energy. The set of all such states forms a continuous manifold of attractor states, related to each other by continuous translation. This manifold can be visualized as the trough of the energy surface, Figure 4A. Rotating a steady state pattern, on the other hand, produces states with higher energy. (Rotation can be visualized as follows. Imagine first cutting open the toroidal periodic network along the edges of the sheet that were originally glued together to produce a periodic network. On the resulting sheet, rotate the pattern, and rejoin the cut edges. This procedure will produce discontinuities in the pattern along the rejoined edges.) Hence the attractor manifold does not include continuous rotations.

Inputs that induce pattern translation will stably move the network state along the trough, even if the inputs are small, and the integrated value of the input will be reflected in the updated network phase. On the other hand, inputs that attempt to induce rotations will not produce lasting changes in network state, because these states are unstable and will quickly (over a few hundred milliseconds or less) decay as the pattern relaxes to its preferred orientation. 
Similarly, distorting the pattern by stretching it, adding noise, or by removing blobs from the pattern will generate an unstable state, which will rapidly decay to a steady state within the attractor manifold.

In the \textit{aperiodic network}, translations of a steady state pattern are similar but not exactly equivalent, because the phase of the activity pattern relative to the boundary affects the energy of the state. Strictly speaking then, these states do not form a continuous attractor manifold, Figure 4B. Instead, the manifold is slightly rippled along the direction of translations. To drive translations, velocity inputs must be large enough to overcome the ripple barrier. This explains why below a critical velocity, the pattern is pinned in our simulations. The ripple amplitude depends on how much influence the boundary has on the network dynamics. If activity fades to zero sufficiently smoothly near the boundary the ripple can be small. Pattern translation then corresponds to motion along a nearly flat direction on the manifold, pinning is confined to a negligibly small range of velocities, and integration of inputs can be accurate. A reduction of pinning can be achieved also by increasing the network size, while keeping the boundary profile fixed, because boundary effects scale as the ratio of network periphery to network area.

A stable population pattern state can be rotated around the center of a circular aperiodic neural sheet to obtain another stable state that is identical in energy to the original one. Hence, rotations correspond to a flat direction in the energy surface, Figure 4B. Any input that couples even slightly with the rotational mode can drive rotations in the network pattern.  The velocity inputs to the network, though configured to drive translational pattern flow, can weakly drive rotations due to boundary effects that couple the translational drive to rotational modes. In spiking networks, discussed below, rotations can be driven also by noise.

In the network models described here, the structure of the attractor manifold (e.g., Figure 4A or 4B) is completely determined by the matrix of pairwise weights between neurons and the inputs received by each neuron. Once the weights between all pairs of neurons and the inputs to each neuron are specified, the matrix does not change if the locations of the neurons on the cortical sheet are shuffled, so long as the weights and inputs to each neuron are held fixed (see \textit{Discussion}). Thus, statements about the existence of a manifold of stable network states and stable SN grid responses, and the predictions that stem from them, do not depend on topography, even when stated here for expositional simplicity in terms of topographically arranged population-level patterns.

\subsection*{Spiking networks and noise}
So far we have considered errors in integration that occur in the absence of noise. Unlike in the noise-free case, neural noise can induce the population pattern to flow or rotate 
even when velocity inputs are absent. To assess how noise influences the precision of the network's response, we present results from spiking neural networks with the same connectivity as in the rate based models. 
Dynamics in these networks are noisy due to the stochasticity of discrete spiking events.

For the same network parameters as in Figure 2, and assuming that neural
firing is an inhomogeneous Poisson process, we find that the periodic network
continues to perform well enough to produce coherent SN responses over long trajectories 
(Figure 5a and Figure S3). In the aperiodic network, performance with Poisson spiking neurons is considerably worse than in the rate based model, enough to destroy the grid-like SN response over a $\sim$130 meter, 10-minute trajectory, in
particular due to rotations (Figure S3). Network performance improves, 
however, if spiking in the network is more regular than implied by inhomogeneous Poisson statistics. To quantify this effect, we performed simulations with sub-Poisson statistics (see \textit{Methods}). The variance of neural firing is characterized, in our simulations, by the coefficient of variation (CV) of the inter-spike interval. 
With a sufficiently low CV, aperiodic network dynamics are precise enough to produce a coherent SN response over a trajectory lasting 10 minutes and $\sim$130 meters, Figure 5b and Figure S3. 

\textit{Quantification of noise-driven translational drift.} Integration can be decomposed into two elements: a memory that holds onto the state of the integrator, and a mechanism that correctly increments the state of the integrator in response to inputs. The linearity of the velocity response of the network, described earlier for noise-free networks, may be viewed as an assessment of the accuracy of the increment mechanism, while the degree of drift in the network state in the absence of velocity inputs and external corrective cues is a quantification of the network's ability to hold onto its current state. 
Therefore, a way to assess the effect of noise on integration accuracy is to examine the drift in the population state when external velocity inputs are absent.

As shown in Figure 6, the states of both periodic and aperiodic spiking networks drift significantly over measurable time-scales, in the absence of any velocity input. As expected, the network state remains in the attractor manifold: Neither network displays stretching or other distortions (data not shown), but the aperiodic network pattern drifts in phase and orientation, while the periodic network pattern drifts in phase without rotation (Figure 6A, B). 

Quantitatively, the drift in the phase of the population pattern appears diffusive (Figure 6C, periodic network): in a time interval $\Delta t$ the square of the average drift due to noise can be written as
\begin{displaymath}
\left< \Delta {\bf x}^2 \right> = D_{\rm trans} \Delta t.
\end{displaymath}
The diffusion constant $D_{\rm trans}$ decreases with network size and increases with the CV of neural spiking (Figure 6D), scaling as  
\begin{displaymath}
D_{\rm trans} \propto \frac{{\rm (CV)}^2}{N}.
\end{displaymath}
This result can be used to obtain an estimate for the maximal expected duration of accurate integration in the presence of noise for networks of different sizes and CVs. Noise can be said to ``decohere'' or destroy the SN response when it drives the network phase to drift by half the pattern period. By this measure, and with the parameters used in Figure 2A-C, we plot in Figure 6E the maximal duration of accurate integration, as a function of
network size, for two values of the CV (1 and 0.5). This duration is about $400$s in a periodic network with 
$10^4$ neurons and  CV$=1$, roughly in agreement with our observations from Figure 5 and Figure S3. 

We recall that in both larger and smaller versions of the deterministic periodic network, integration was highly accurate, Figure 2A-C and Figure S1. The relatively weak dependence on network size in the deterministic case gives way to a stronger sensitivity on size in the presence of neural noise: the interval of accurate integration, set by noise-driven drift, decreases linearly with decreasing network size. Thus, neural noise sets limits on the minimum size of the network needed to produce accurate integration, even in the periodic network. 

\textit{Quantification of noise-driven rotational drift.}
In aperiodic networks, rotational drift of the population pattern can be measured by tracking the orientation of the pattern as a function of time. We find that this drift too is diffusive:
\begin{displaymath}
\left<\Delta \theta^2\right> = D_{\rm rot} \Delta t.
\end{displaymath}
The diffusion constant can be measured in a similar fashion to the measurement of $D_{\rm trans}$ in Figure 6C. Roughly, $D_{\rm rot} \propto {\rm (CV)}^2$, Figure 6F. We can use these measurements to obtain an estimate for the maximal expected time until noise-driven rotations destroy the single neuron pattern during path integration: Requiring that the rotational drift remain smaller than $\pi/12$, we obtain an estimate of about $85\,$s for a network with CV$=1$, and about $680$\,s for CV=$1/\sqrt{8}$, in agreement with the time over which accurate integration was observed in Figure~5 and in Figure~S3. 

Assuming that the translational drift in the aperiodic network is similar to that measured in the periodic network we conclude that, in the aperiodic network, rotations are the more severe source of noise-driven decoherence of the SN response. This conclusion is in agreement with the observation that the $128^2$ aperiodic network required a smaller CV, compared to the periodic network (where there are no rotations) to achieve a similar performance, even though the two networks showed similar performance in the noise-free case. 

\textit{Variability in recorded grid cell responses.} 
Motivated by the result that sub-Poisson spiking statistics are important for accurate integration in the
grid-cell network, we analyzed spike recordings from neurons in dMEC \citep{Hafting05}. Under certain conditions, cortical neurons are reported to be Poisson or even super-Poisson in their firing statistics \citep{Softky93,Shadlen94}. Interestingly, our analysis of the dMEC data suggests that grid cell firing is significantly sub-Poisson (Figure S4).

For various reasons, it is not possible to exactly compare the CV used in our simulations and the CV of the recorded cells in dMEC. For example, dMEC contains numerous cell types, each of which may have different CVs. Also, the effects of individual neural variability on integration performance are ameliorated by averaging over the network population, but the size of the actual dMEC network may not be the same as in our simulations, and the actual network may contain correlations not included in our model, so that even if we were able to pick the ``correct'' CV for individual neurons, the net effect on integration performance may be different in the model from that in dMEC. Finally, the CV is a low-dimensional measure that does not fully characterize the spiking statistics of a neuron: even if we could match the size of the dMEC network and the CV of each neuron type, the statistics of our model neurons could greatly differ from those in the rat. 

Despite these caveats, our results suggest that a significant blurring of the SN response is expected to occur on a time scale ranging between a few minutes to a few tens of minutes, within a reasonable range of estimates for the number of neurons in the network and the variability of neural spiking.

\section*{Predictions of the attractor model}

Armed with the proof-of-concept results that a continuous attractor network model can integrate velocity inputs accurately enough to produce SN grids, we next seek to explore testable predictions of the continuous attractor hypothesis in the grid cell system and contrast them with the properties of models in which the grid responses emerge independently in each cell \citep{OKeefe05,Burgess07,Hasselmo07}. Unless explicitly specified, all proposed tests are intended for conditions in which external, spatially informative cues have been removed. 

{\bf Stability of the attractor manifold. \ } As described earlier, the low-dimensional structure of the attractor means that only a very small subset of possible states of the network, defined by strict inter-relationships in neural activity (population patterns), are stable, while other states quickly decay away. The quantity conserved across pattern translations and therefore across the attractor manifold 
is the phase {\em relationship} between cells, defined by whether neurons are co-active or active at different phases. The stability of the attractor manifold and the instability of states outside it have a number of implications for experiment.

\textit{Stability of phase relationships in absence of inputs. \ } Due to the stability of the attractor manifold, phase relationships in the periodic network should be stable over the time-scale of days (because the pattern does not rotate), regardless of inevitable drifts in the absolute phase of individual neurons. Even in aperiodic networks, we expect phase relationships to persist over 1-10 minutes, but possibly not longer due to the possibility of rotations. Under similar conditions in models where the grid is generated separately by individual neurons (``independent neuron models''), like temporal interference models \citep{Burgess07,Hasselmo07}, the phase relationships between cells should drift or random walk over relatively short periods of time, on the same time-scale as drifts in the absolute phase of single cells. This is because in independent neuron models, the phase of the grid response of each cell is determined individually, in part from the phase of an intrinsic oscillator. Hence, unlike the continuous attractor models, phases of different neurons are untethered to each other through network interactions. 

\textit{Stability against small perturbations of neural subsets. \ } Because the attractor dynamics are restoring, small perturbations (small induced changes in the activity of neurons) of state without a component along the attractor manifold should not produce lasting changes in the states of these neurons or the network. Network interactions should restore the state to the original state that preceded the perturbation: thus, both the absolute phases of cells and their phase relationships should be unchanged by the perturbation. This statement also applies to large perturbations, if they have no appreciable projection along the attractor manifold (e.g., large random perturbations made directly to different layer II/III grid cells with low velocity sensitivity are examples of such large perturbations). By contrast, following small or large perturbations to subsets of cells in independent neuron models, the absolute activity states of those cells, as well as their relative phase relationships with unperturbed neurons should change, due to the absence of restoring network interactions. 

\textit{Coherent movement along the attractor manifold in response to incoherent perturbations. \ }
Perturbations that have a large component along the attractor manifold should drive a coherent transition to the point on the attractor manifold that is closest to the perturbed state. Because the new state will be on the attractor manifold, phase relationships between neurons should be unchanged. Head direction cells provide a means to induce such a perturbation: Stimulating a subset of head direction  cells should drive a rigid (coherent) and lasting translation of the entire population pattern, producing the same shift in phase in all cells, regardless of whether or not they received direct head direction input. By contrast, similar inputs provided only to subsets of cells in independent neuron models should produce changes in phase only in the stimulated cells.

{\bf Single neuron responses. \ } The continuous attractor model predicts that all cells in the network must have identical orientations, and all phases must be equally represented in the population \citep{Fuhs06}. Both these properties are consistent with observations \citep{Hafting05}, but are difficult to explain in independent neuron models, without invoking additional mechanisms that effectively turn the system into a low-dimensional attractor.

Further, in the continuous attractor model, if any cell's grid response contains a reproducible irregularity of any kind (e.g., a global skewing of the lattice, or a local defect, such as a local 5-7 pairing of neighbors instead of the usual 6), it follows that {\em every} cell in the network must display the same irregularity, up to a global shift in phase. Indeed, our preliminary analysis of data from \citep{Hafting05} supports this prediction, Figure S5.

{\bf Expansion or contraction of the SN grid in different environments. \ } In experiments where a familiar enclosure is resized, the SN response is observed to rescale along the rescaled dimension of the enclosure, at least temporarily \citep{Barry07}. Further, when the rat is placed in a novel environment, the SN grid responses are observed to isotropically expand or contract \citep{Fyhn07}. These observations have sometimes been interpreted as evidence against the continuous attractor models of grid cells.

To explain why these rescaling experiments are consistent with a continuous attractor model of grid cells, it is important to stress the difference between the population-level and the SN responses. The attractor manifold consists of the steady states of the population response, which consists of translations (and in aperiodic networks, rotations) of a canonical pattern. Thus, stretching and rotation of the population pattern are forbidden (unstable) and cannot be invoked within the continuous attractor models to explain the experimental observations. 

The SN response, on the other hand, 
is not directly subject to constraints imposed by the attractor manifold on the population pattern, 
because it is a function of both the instantaneous population pattern and the velocity response of the pattern in time. If the pattern were to flow more slowly along one dimension than the other, for equivalent rat speeds, the SN response would be a stretched version of the regular underlying population grid, with the stretched dimension corresponding to the slow flow dimension. Hence, stretching of the SN response can be explained in the continuous attractor model by an amplitude modulation of head direction inputs tuned to the relevant head direction, without inflicting such a deformation on the population pattern (Figure 7 A--B). If the population pattern were not constrained by the low-dimensional attractor, SN stretching could instead be effected by a stretching of the population pattern in the cortical sheet, Figure 7B (rightmost column). 

How can experiments distinguish between these two possibilities? The continuous attractor model predicts that  the phase relationships between neurons must remain unchanged upon stretching of the SN response (Figure 7 A-B, middle column). This prediction of the continuous attractor model will be explicitly violated if stretching happens at the population level, Figure 7 A-B, rightmost column. Further, the continuous attractor model predicts that the strength of velocity modulation in the head direction inputs to dMEC and in the conjunctive heading- and velocity-sensitive grid cells \citep{Sargolini06} should decrease along the grid's stretched dimension, which corresponds to the expanded enclosure dimension, and the percentage decrease should correspond exactly to the percentage stretching of grid responses. 

In contrast, if the SN stretching is due to a similar stretching in the population response, there should be little to no change in the amplitude of velocity modulation of the cells. In summary, changes in the phase relationships between cells, or no change in the velocity modulation of the head direction inputs to dMEC, when the SN responses have been stretched, would be evidence against the attractor model. 

Similarly, a rotation \citep{Hafting05} (or an isotropic stretching \citep{Fyhn07}) of the SN response, which happens when the cue-card is rotated (or when the enclosure is novel), is predicted to be due to an isotropic rotation (or scaling in the velocity-modulated amplitude) of the head direction inputs to the network, while the network pattern is predicted to remain unrotated (unstretched), Figure 7 A,C. The former part of the prediction, about the rotation of head direction inputs to the grid cell network, is consistent with separately observed responses in head direction cells to cue card rotations \citep{Taube90,Taube95}.

{\bf Insufficiency of feedforward input and necessity of recurrent processing for spatial periodicity. \ } Lidocaine, or another blocker of spiking activity, applied locally to dMEC without affecting inputs to dMEC should abolish periodic spatial responsiveness in the subthreshold activity of grid cells. This is because all periodic patterning in the continuous attractor model arises from recurrent interactions within dMEC. By contrast, individual-neuron models, where the computation is performed within each neuron, may continue to show spatially periodic responses under such a manipulation. 


{\bf Distinguishing between attractor models. \ } Given that both periodic and aperiodic continuous attractor network models of dMEC are capable of accurate integration of rat velocity inputs, how might it be possible to experimentally distinguish between the two possibilities?

A periodic network shows no pinning, and rotations of the population response are forbidden. Thus, phase relationships between neurons should be absolutely stable over very long times even in the absence of any sensory inputs. By contrast, aperiodic networks should be pinned for sufficiently low velocity inputs, and in the absence of external corrective cues, are expected to rotate on slow timescales (minutes to 10's of minutes). A population-wide rotation will be manifest in altered phase relationships between single neurons, or it could be probed by looking at differential (relative) rotations in the orientation of quickly estimated SN grids versus the head direction cell population.

Next, in an aperiodic network, neurons at the boundaries must receive fading input, meaning that their maximal activity is substantially lower than that of neurons in the bulk; thus, the distribution of maximal rates across grid cells of the same type in an aperiodic network should be wide. If the maximal firing rate of every cell (of the same type) in the network is roughly the same, it would be inconsistent with an aperiodic network. The converse need not be true (i.e., a wide distribution of cells does not imply an aperiodic network, or rule out a periodic network). 

We emphasize that the boundaries of the neural population are not related to physical boundaries in space. Hence the neurons at the boundaries, discussed above, are not expected to bear a relationship to the recently discovered cells in dMEC whose receptive field encodes the rat's proximity to boundaries in the environment \citep{Savelli08,Solstad08}.

Finally, if defects exist in the single neuron response, they may help distinguish between a periodic and an aperiodic network. By defects, here we only mean those arising spontaneously from the pattern formation process in a network whose connectivity is itself defect-free. Defects arising from imperfections in the weights will not flow in response to velocity inputs, and are therefore not expected to produce a systematic defect in the SN response. In the aperiodic case, any defect in the SN response must be eliminated if the rat returns to the area where the defect was observed after first moving in one direction until the defect has flowed off the population pattern. Conversely, if the defect persists upon return to the vicinity of the defect location even after long excursions, the lattice has periodic boundaries. The presence of a stable defect which is present in {\em all} SN responses would incidentally be strong evidence of a continuous attractor network.

The last two predictions can help to distinguish even a well-tuned aperiodic network, which may show relatively little rotation or pinning, from a periodic network.

\section*{Discussion}

The three main contributions of this work are:

(1) A demonstration through modeling that under reasonable conditions grid cells can be good velocity integrators, and more specifically, that continuous attractor models are capable of accurate path integration. 

By `good' integration, we mean that if the model network is given accurate velocity inputs, it produces an accurate estimate of rat position over comparable distance and time-scales to those probed in behavioral assays. Within a plausible range of estimates for network size and neural stochasticity, higher accuracy was reached in larger and relatively noise-free networks, sufficient to reproduce coherent grid cell patterns in response to the full trajectories from \citep{Hafting05}, lasting 10-20 minutes. Smaller networks with more stochastic dynamics were capable of good integration over smaller paths, still consistent with behavioral constraints. 

(2) Furnishing good upper bounds on idiothetic path integration accuracy within dMEC. 

A notable finding is that even noise-free, large networks (periodic and aperiodic) have only finite integration accuracy, and this level of accuracy is only a factor of 10-100 larger than known behavioral abilities. We provide estimates of integration accuracy in the presence of neural noise, which are in the range of 1-10 minutes. Integration performance in a fixed-size periodic network is not expected to vary greatly with parameter tuning; aperiodic networks are more sensitive to parameter tuning, and we have not optimized all parameters. However, aperiodic networks 
are upper-bounded in their performance by the corresponding periodic network. Thus, we expect our estimates to serve as reasonable upper bounds on integration accuracy in dMEC, within the continuous-attractor picture.   

(3) Providing predictions that can falsify the continuous attractor hypothesis and help distinguish between the possibilities that grid responses are generated through continuous attractor networks or through independent cell computations.   

So far, the predictions of continuous attractor models are consistent with the full corpus of grid cell data, and explanatory of many results from experiment, suggesting, when combined with conclusion (1), that continuous attractor dynamics are a viable, relevant mechanism for grid cell activity and path integration. 

{\bf Assumptions of the model. \ } 
Accurate behavioral dead reckoning is a cascaded result of accurate velocity input (relative to the rat's motion) and accurate integration of that input. Our interest in this work was in assessing how well continuous attractor models of dMEC can integrate their inputs. Thus, we did not focus on potential inaccuracies (noise or biases) in the velocity inputs themselves. Even if the network were a perfect integrator, errors in the input would produce an incorrect position estimate. Such errors are likely to play a role in reducing the behavioral range over which rats display accurate dead-reckoning. 

A strength of attractor networks is that responses are self-averaging over the full network: if the velocity inputs are unbiased estimators of rat movements, but are noisy, or if the velocity inputs to the network are not perfectly balanced in number for all directions, the full network will average all its inputs, and the net pattern flow will only reflect this average. For accurate position estimation, however, it is important and therefore likely that inputs to the network are well tuned. 

Another factor that could degrade integration performance is inhomogeneity or stochasticity in the recurrent network weights. While stochasticity in neural activity causes the network state to drift along the attractor manifold, variability in network connectivity modifies the structure of the attractor manifold itself. If recurrent connectivity deviates significantly from the translation-invariant form needed to ensure that all translations of the pattern are accessible without crossing over energy barriers, the activity pattern can become pinned at particular phases \citep{Ernst01}, reducing the fidelity of the network response to small velocity inputs. 

Because knowledge about synaptic strengths in the brain is exceedingly limited, it is unclear what level of variability should be expected in dMEC weights, and whether this amount is sufficient to cause significant pinning. A question for theory, not addressed in this work, is to estimate the amount of variability in the network weights that would be sufficient to reduce the accuracy of integration below that observed in dead reckoning behavioral experiments. For experiments, the difficult challenge is to obtain an estimate of variability in dMEC connectivity.

{\bf Network size. \ } The network size estimate in our continuous attractor model ($10^3-10^4$ neurons) may be viewed as a wasteful proposed use of neurons, but it is broadly consistent with estimates for the total number of neurons in the entorhinal cortex \citep{Amaral90, Mulders97, Augustinack05}. By contrast, independent neuron models \citep{Burgess07,Hasselmo07,Giocomo07}, which do not require populations of neurons to produce grid cell responses, make far more parsimonious use of neurons. In such models, a natural question is to understand what function may be served by the large number of neurons in dMEC. 

Within dMEC, the breakdown of total neural allocation, between neurons per grid network versus the number of different grid networks, is unknown. dMEC might consist of a very large number of very small networks with different grid periods, which is optimal for representational capacity \citep{Fiete08}. (For a fixed neuron pool size, the addition of neurons per grid at the expense of the total number of different grids causes a large capacity loss \citep{Fiete08}.) But the dynamical considerations presented here suggest otherwise, because accurate path integration in each grid requires many neurons. In contradiction to optimal capacity considerations, therefore, continuous attractor models predict a large membership in each grid network, and correspondingly few different grids.

A fascinating question is whether the discrete islands of cells observed in anatomical and imaging studies of cells in layer II of the human and primate entorhinal cortex
\citep{Hevner92,Goldenberg95,Solodkin96,Augustinack05, Witter06}, 
as well as indications in rodents for modular structure in dMEC
\citep{Dickson00,Witter06} 
correspond to separate attractor networks, in which case the number of different grid periods can be directly inferred.

{\bf Periodic versus aperiodic networks. \ } We have shown that both periodic and aperiodic networks can perform accurate integration. Which topology is dMEC likely to posses? The models and results of this work are largely agnostic on this question.
However, the aperiodic network requires fine-tuning of its parameters to perform nearly as well as an untuned periodic network. Even after fine-tuning, integration in the periodic network tends to be better, because unlike in the aperiodic case, the population pattern cannot rotate. Thus, from a functional perspective, periodic boundaries are preferable over aperiodic ones.

Other constraints on network topology may stem from the developmental mechanism of the grid-cell network. Such developmental constraints could overrule potential functional preferences, in determining network topology. 

{\bf Network topography. \ } If neural locations in the cortical sheet are scrambled, while preserving the neural indices $i$ and the pairwise weights $W_{ij}$ between neurons, the grid-like patterning in the cortical sheet will disappear, but there will be no change in the single neuron triangular lattice response or in any other dynamical property of the network. The underlying structure of the attractor manifold (e.g., whether or not it is continuous) is a function of network connectivity, but does not depend on the layout of neurons on the cortical sheet. Thus, the lack of topography observed in experiments, in which neighboring neurons have different phases, is not a problem for the dynamics of continuous attractor models of grid cell activity. Instead, the problem is one of learning: how does a network wire up so that the intrinsic structure of the weight matrix resembles center-surround connectivity, but the neurons are themselves not arranged topographically in space?

{\bf The problem of learning.\ }
A topographic, aperiodic model network would have relatively simple wiring rules (if we ignore the directional neural labels and corresponding segregation of head-direction inputs and shifts in the outgoing weights required for the velocity-coupling mechanism): each neuron would simply have spatially restricted center-surround interactions with its neighbors. This has prompted the observation that such a topographic network could serve as a starting point for the development of a network with a less topographical layout and  periodic boundaries \citep{McNaughton06}. For instance, the proposal by \citep{McNaughton06} for wiring an atopographic and periodic network is based on three assumptions: (1) that another area, the `teacher', contains an initial aperiodic, topographic network with population grid patterning and no velocity shift mechanism, (2) that the network pattern, when subject to intrinsic or extrinsic noise, tends to translate without rotation, (3) that the network projects through spatially random connectivity to the naive dMEC, and activity-dependent activity mechanisms within dMEC cause neurons that are coactivated by the teacher network, to wire together. However, results from the present work show that the fundamental features of aperiodic networks pose a problem for such a scheme.

We showed that the population pattern in a deterministic aperiodic network fully equipped with a translational velocity shift mechanism and driven by purely translational velocity inputs, tends to rotate within a few minutes. This is the short end of the time-scales over which plasticity mechanisms for network development would act. If the network is entirely driven by noise and lacks a specific velocity shift mechanism (as in \citep{McNaughton06}), the problem is far worse: undesirable rotations become as likely as translations, and the pattern orientation can decohere in seconds, invalidating assumption (2). Thus, the precursor network pattern will not be able to entrain a periodic grid in the target network. 

The problem of pattern rotations over the time scale of learning is pertinent for any effort to produce a periodic network from an initially aperiodic one in the absence of anchoring sensory inputs and a velocity coupling mechanism.  

{\bf The elusive hypothesis. \ } The concept of low-dimensional continuous attractors has influenced our understanding of neural systems and produced successful models of a number of neural integrators \citep{Skaggs95, Seung96, Zhang96, Goodridge00, Seung00, Xie02, Stringer03}. Yet proof of continuous attractor dynamics
(or some discrete approximation to continuous attractor dynamics) 
in the brain has remained elusive: experiments in supposed continuous attractor systems have failed to unearth evidence to conclusively validate or falsify the continuous attractor hypothesis. The relative richness (e.g., size, dimensionality of the manifold) of the grid cell response compared to other possible continuous attractor systems may provide a more structured and unambiguous testing ground for predictions stemming from the continuous attractor hypothesis. 
Testing of these predictions, many based on cell-cell correlations, is feasible with existing experimental technologies, and such tests may help to determine whether a low-dimensional continuous attractor is central to the dynamics of the grid cell system. 

\section*{Methods}

The dynamics of rate-based neurons is specified by:
\begin{equation}\label{eq:generaldynamics}
\tau \frac{d s_i}{dt} + s_i = f\biggl[\sum _{j}W_{ij} s_j +B_i\biggr]
\end{equation}

The neural transfer function $f$ is a simple rectification nonlinearity: $f(x)=x$ for $x>0$, and is 0 otherwise. The synaptic activation of neuron $i$ is $s_i$; $W_{ij}$ is the synaptic weight from neuron $j$ to neuron $i$. The time-constant of neural response is $\tau=10$ ms. The time-step for numerical integration is $dt=0.5$ ms.

We assume that neurons are arranged in a 2-d sheet. Neuron $i$ is located at ${\bf x}_i$. There are $N=n\times n$ neurons in the network, so $x$ ranges from $(\frac{-n}{2},\frac{-n}{2})$ to $(\frac n 2, \frac n 2)$. We use $N=128^2$ in all figures except where specifically indicated. Each neuron $i$ also has a preferred direction (W, N, S, E) designated by $\theta_i$. Locally, each $2\times 2$ block on the sheet contains one neuron of each preferred direction, tiled uniformly.

The preferred directions are restricted to N,S,E,W for convenience in modeling; in the rat, these preferences might span the continuum $[0,2\pi]$. The preferred orientation of a neuron is used to (1) determine the direction in which its outgoing weighs are shifted, and (2) determine the rat velocity inputs it receives.

The recurrent weight matrix is
\begin{equation}\label{eq:weights1}
W_{ij}=W_0({\bf x}_i-{\bf x}_j- l ~ \hat{\bf e}_{\theta_j})
\end{equation}
with
\begin{equation}
W_0(\bf x) = a ~ e^{-\gamma |\bf x|^2}-e^{-\beta |\bf x|^2}
\end{equation}
The weight matrix has a center-surround shape, but is centered at the shifted location ${\bf x}-l ~ \hat{\bf e}_{\theta_j}$. Implicit in the form of the weight matrix, where connectivity is a function of neural separation, is the assumption that neurons are topographically arranged. This is not a necessary requirement (see  \textit{Discussion}), but does greatly facilitate visualization and presentation. In all simulations, we used $a=1$, $\gamma =1.05 \times \beta$, and $\beta =3/\lambda_{\rm net}^2$ where $\lambda_{\rm net}=13$ is approximately the periodicity of the formed lattice in the neural sheet. With $a=1$, all connectivity is inhibitory; thus, local surround inhibition alone is sufficient to reproduce gird cell responses, but the network could include excitatory interactions ($a>1$) without qualitatively affecting the results.

The feedforward input to neuron $i$ is
\begin{equation}\label{eq:inputs}
B_i=A(x_i)~ \bigl(1+\alpha ~ \hat{{\bf e}}_{\theta_i} \cdot {\bf v} \bigr)
\end{equation}
where $\hat{{\bf e}}_{{\theta}_i}$ is the unit vector pointing along $\theta_i$, and ${\bf v}$ is the velocity vector of the rat, measured in m/s. If $l=0$ (Eq. \ref{eq:weights1}) and $\alpha=0$ (Eq. \ref{eq:inputs}), the network generates a static triangular lattice pattern, Figure 1A, with overall intensity modulated by the envelope function $A$ (e.g., Figures 2D, 3B, and 3 D1-D4).

If $l,\alpha$ are non-zero, they allow rat velocity (${\bf v}$) to couple to the network dynamics, and drive a flow of the formed pattern. The magnitudes of both $l$ and $\alpha$ multiplicatively determine how strongly velocity inputs drive the pattern, and thus control the speed of the flow of the pattern for a fixed rat speed. The triangular lattice pattern is only stable for small values of the shift $l$ in the outgoing weights, thus we keep $l$ fixed so that the outgoing weights are shifted 2 neurons. With $l$ fixed, $\alpha$ determines the gain of the velocity response of the network. If $\alpha |{\bf v}|\ll 1$, we can expect the velocity inputs to drive pattern flow without destroying the stability of the formed lattice.
In the plots shown, $\alpha=0.10315$.
The grid spacing of the SN response is ultimately determined by two factors: (i) The grid spacing of the population response, which is set by the shape of the symmetric weight matrix $W_0$, and (ii) the gain of the network's flow response to a velocity input, which depends on $l$ and $\alpha$.

The envelope function $A$ spatially modulates the strength of the inputs to the neurons, and can scale neural activity without disrupting the lattice pattern. This can be seen from Equation \ref{eq:generaldynamics}: if the input $B$ is uniform, then scaling $B$ is equivalent to scaling $s$. It is important to observe that the velocity inputs must also be modulated by the envelope $A$, Eq.~\ref{eq:inputs}, to insure the same flow rate in the faded regions as in the bulk. This is because the local flow rate is given by the velocity-modulated component of the feedforward input divided by the total feedforward input.

For the network with periodic boundary conditions, the envelope function is 1 everywhere. For the aperiodic network,
\begin{equation}
A({\bf x}) = \left\{
\begin{array}{cl}
1 &  \ \ \ \ |{\bf x}| < R-\Delta r \\
{\rm exp}\left[-a_0 \left(
\frac{|{\bf x}|-R+\Delta r}{\Delta r}
\right)^2 \right]& \ \ \ \ R - \Delta r \le |{\bf x}| \le R
\end{array}
\right.
\end{equation}
$R$ is the diameter of the network and $a_0=4$ (for example, see Figure 2D and Figure 3B, D1-D4). In Figure 3 (D4), R = 128; in all other figures, R = 64. 
The parameter
$\Delta r$ determines the range of radii over which input tapering occurs: The larger $\Delta r$, the more gradual the tapering. In all the aperiodic simulations $\Delta r = R$, except for Figure 3 (A-C and D2, D4), where $\Delta r = 32$ and Figure 3 (D3), where $\Delta r = 16$.

{\bf Spiking simulations:}
To simulate a Poisson process (CV=1, where CV is the ratio of the inter-spike interval standard deviation with the mean), in each time-step $[t,t+\Delta t]$ neuron $i$ spikes with probability given by $P_{\rm spk}(i; t, t+\Delta t) = f(W_{ij}s_j(t)+B_i(t)) \Delta t$ (in our simulations, $f_i$ is always much less than $1/\Delta t=200$, ensuring that $P_{\rm spk} \ll 1$). The synaptic activation $s_i(t)$ is computed from neural spiking: it increments by 1 at time $t$ if neuron $i$ spiked at $t$, and otherwise decays according to

\begin{equation}
\tau \frac{ds_i}{dt} = - s_i(t)
\end{equation}
The process for generating spike trains with CV$=1/\sqrt{m}$ (for integer-valued $m$) is similar to that for generating a Poisson train. 
We first subdivide each interval into $m$ sub-intervals of length $\Delta t/m$ each, and simulate on this finer time resolution a fast Poisson spiking process with rate $m*f(W_{ij}s_j(t)+B_i(t))$. We then decimate the fast Poisson process,  retaining every $m$-th spike and discarding all the other spikes. This procedure generates a spike train with rate $f(W_{ij}s_j(t)+B_i(t))$ and CV$=1/\sqrt{m}$.

{\bf Initial conditions} Aperiodic network: initially network activity is low; neurons receive external input with ${\bf v}=0$ in addition to a small independent random drive, which leads to spontaneous pattern formation. Periodic network: we initialize an aperiodic network with otherwise identical parameters, and after pattern formation apply periodic boundary conditions. The parameters for the aperiodic network have to be chosen to be commensurate with the size of the network to avoid excess strain and the formation of defects when the boundaries are made periodic. We flow both the periodic and aperiodic network states with unidirectional velocity inputs, corresponding to a velocity of 0.8 m/s, in three different directions ($0, \pi/5, \pi/2-\pi/5$) for 250 ms each to heal any strain and defects in the formed pattern. After this healing period, we give as input to the network either real rat velocity (data obtained by differentiating recorded rat trajectories -- published in \citep{Hafting05} -- then linearly interpolating between the recording time-steps and the time-step $dt$ in our simulations), or a sequence of velocity steps (described next).

{\bf Velocity response curves:} The network is initialized to the exact same initial template state at the beginning of each step (using a template pattern stored following one run of the initialization process described above). Each step consists of a constant velocity input, with one of four directions ($0$, $\pi/6$, $\pi/3$, $\pi/2$). The velocity is incremented in steps of 0.02 m/s. We use only the second half of the 5 s long steps to compute the network's velocity response.

{\bf Tracking lattice orientation and flow}: We track how far the pattern has flowed beyond a lattice period and beyond the scale of the network by continuously recording the velocity of the blob closest to the center, and integrating the obtained velocity. We track the orientation of the lattice by computing its Fourier transform and recording the angles of the three blobs closest to the origin in Fourier space.

To assign units of centimeters to the accumulated network pattern flow and compare it to rat position (Figure 2\,C, Figure F, 3\,C, Figure S1, and Figure S3),
 we must obtain the scale factor relating the network pattern flow velocity to the velocity of the rat. The scale is determined by optimizing the match between network flow velocity and the derivative of the rat position throughout the simulation. The offset is set so that the network drift at time $t = 0$ is zero.

\section{Acknowledgments}
We are grateful to Mehran Kardar
and Michael Cross for helpful conversations. 


\pagebreak

\renewcommand{\figurename}{Figure}

\centerline{
\includegraphics[width=0.7 \textwidth]{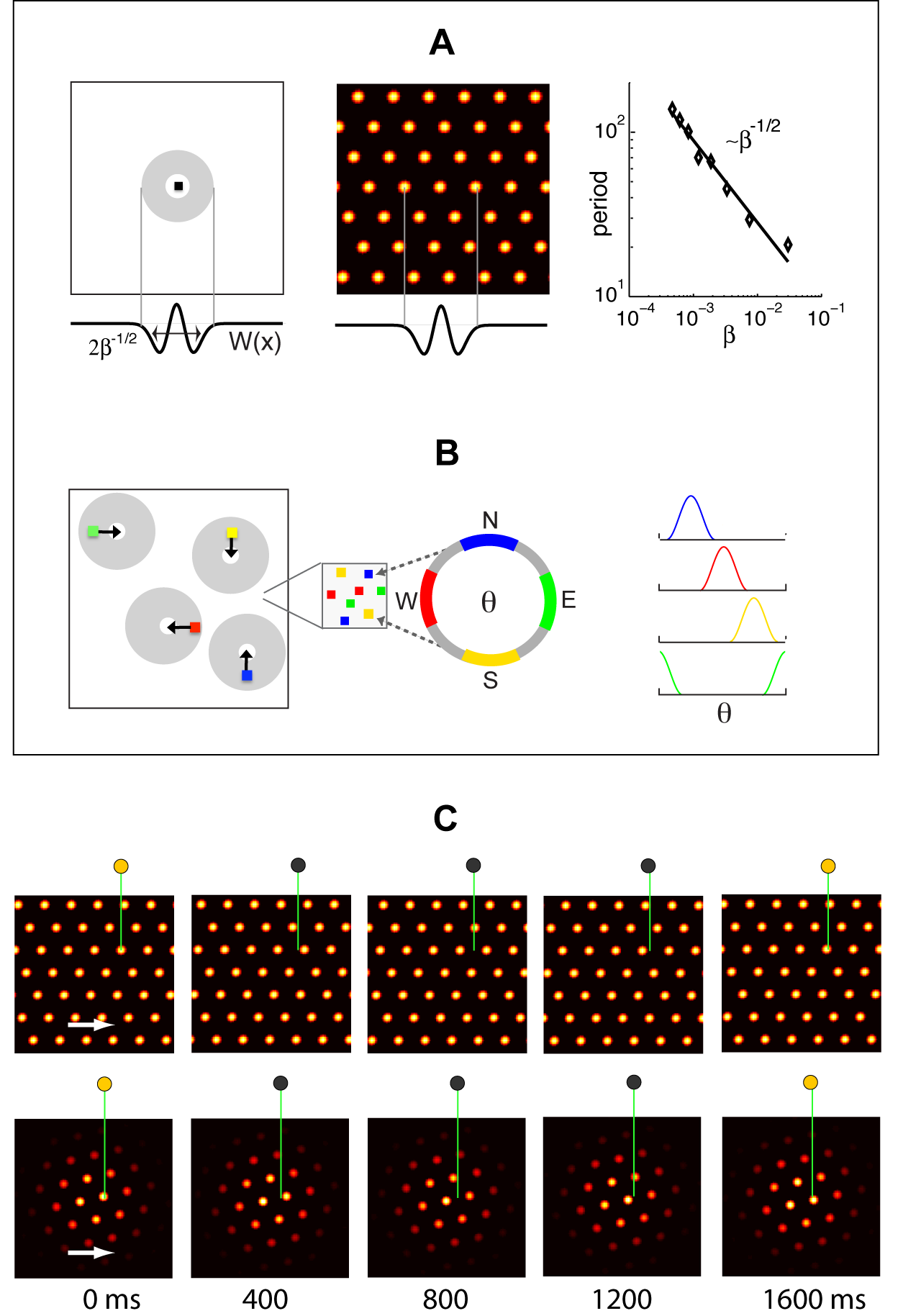}}

\begin{figure}[tbh] \label{fig:methods}
\caption{\textbf{Network architecture and response.} {\bf A} Pattern formation in the neural population: Left, schematic depiction of the outgoing weights of a neuron in the network. All neurons have the same connectivity pattern, and the width of the inhibitory surround is parameterized in our model by $\beta^{-1/2}$ (see \textit{Methods}). Center, circularly symmetric center-surround connectivity, with sufficiently strong local inhibitory flanks, produces a regular triangular lattice population pattern in the neural sheet through spontaneous destabilization of the uniform mode (Turing instability). Right, the pattern period depends on the width of the inhibitory surround.
{\bf B} The velocity shift mechanism by which velocity inputs drive pattern flow: Each neuron in the sheet is assigned a preferred angle (color coded), which means two things. First, the outgoing weight profile, instead of being centered exactly on the originating neuron, is shifted by a small amount along the preferred angle in the neural sheet (left). Each patch in the neural sheet contains neurons with all preferred angles. Second, the direction preference means that the neuron receives input from head direction cells tuned to the corresponding angle (center and right). 
{\bf C} Snapshots of the population activity, when the networks (periodic boundaries, above; aperiodic boundaries, below) are driven by a constant velocity input in the rightward direction. In the periodic network, as the pattern flows, it wraps around the opposite edge. In the aperiodic network, as the pattern flows, blobs move away from the left edge and new ones spontaneously form through the same dynamics that govern pattern formation. (Boundaries are considered in more detail in the paper and in later figures.) The green lines represent an electrode at a fixed location in the neural sheet, and the circle above them represents the activity state of the targeted neuron (gray=inactive, yellow=active). Network parameters are as in Figure 2\, A--C and Figure 2\,D--F.
}
\end{figure}


\begin{figure*}[bth]
\centerline{
\includegraphics[width=0.95 \textwidth]{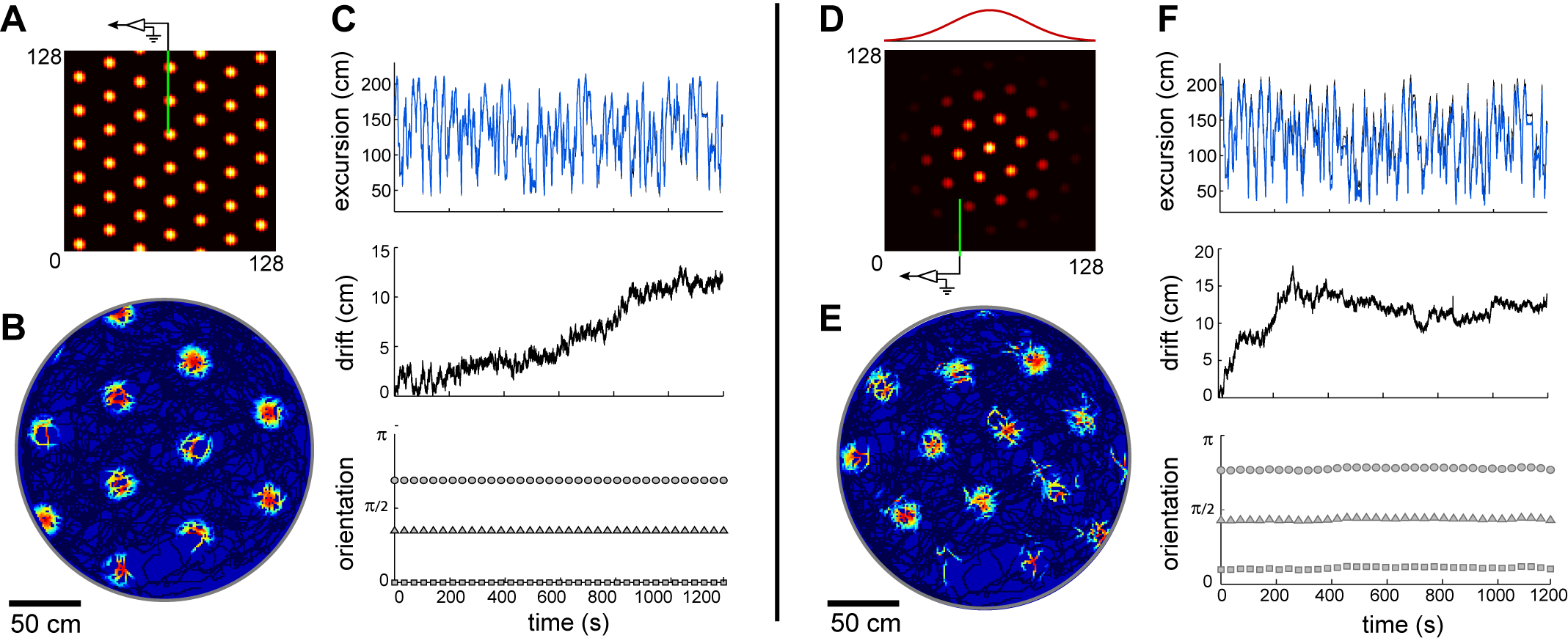}}

\caption{\textbf{Periodic and aperiodic networks are capable of accurate path integration.} Simulation of network response, with velocity inputs corresponding to a rat's recorded trajectory in a 2m circular enclosure \citep{Haftingdatarelease}. The boundary conditions in the neural sheet are periodic in panels A--C and aperiodic in panels D--F. {\bf A}, {\bf D} Instantaneous activity within the neural sheet (color represents the firing rate: black corresponds to vanishing rate). The red curve in D
represents the fading profile of inputs to the network. {\bf B}, {\bf E} Grid cell response: average firing rate of a single neuron (located at the electrode tip in panels A, D), as a function of the rat's
position within the enclosure. {\bf C}, {\bf F} Velocity integration in the network: Top: Actual distance of the rat from a fixed reference point (black), compared to the network's integrated position estimate, obtained by tracking the flow of the pattern in the population response (blue). The reference point is at the left-bottom corner of the square in which the circular enclosure is inscribed. Middle: Accumulated distance
 between the integrated position estimate and the actual position. Bottom: 
Orientation of the three main axes in the population response during the trajectory. Note that there is no rotation in the periodic network, and little rotation in the aperiodic one.}
\end{figure*}


\begin{figure*}[tbh]
\centerline{
\includegraphics[width=0.94 \textwidth]{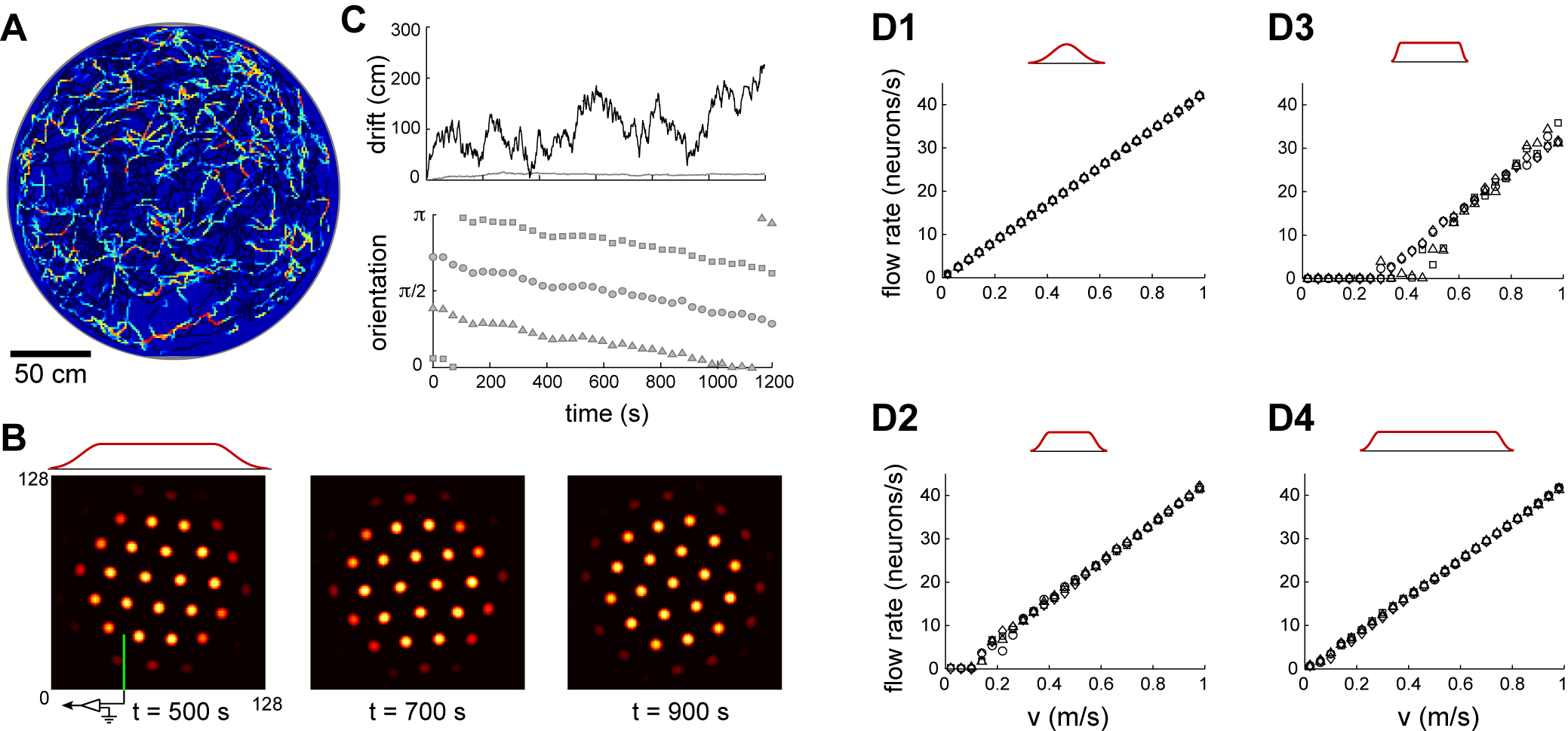}}

\caption{\textbf{Boundary conditions and network size strongly affect fidelity of network response.} {\bf A}--{\bf C} Same simulation as in Figure 2D--F, but with a sharper input profile (red curve above B). The SN pattern has no periodicity (A), the integration error is large (thick line in C, upper plot; note the different scale compared to Figure ~2\,E, whose error is represented by the thin line), and the population response rotates frequently (C, lower plot). {\bf D1}--{\bf D3} Network velocity response as a function of different input profiles: Input profile decay is least abrupt in D1, more abrupt in D2, and most abrupt in D3 ($\Delta r = 64, 32,16$ for D1, D2, and D3, respectively; network size is 128 neurons per side($R=64$) for all). {\bf D4} The input profile at the boundaries is identical to D2 ($\Delta r = 32$), but the network is larger (256 neurons per side or $R = 128$). Panel D2 corresponds to the parameters in A--C, and panel D1 corresponds to the parameters in Figure 2\,D--F. }
\end{figure*}


\begin{figure}[bth]
\centerline{
\includegraphics[width=0.47 \textwidth]{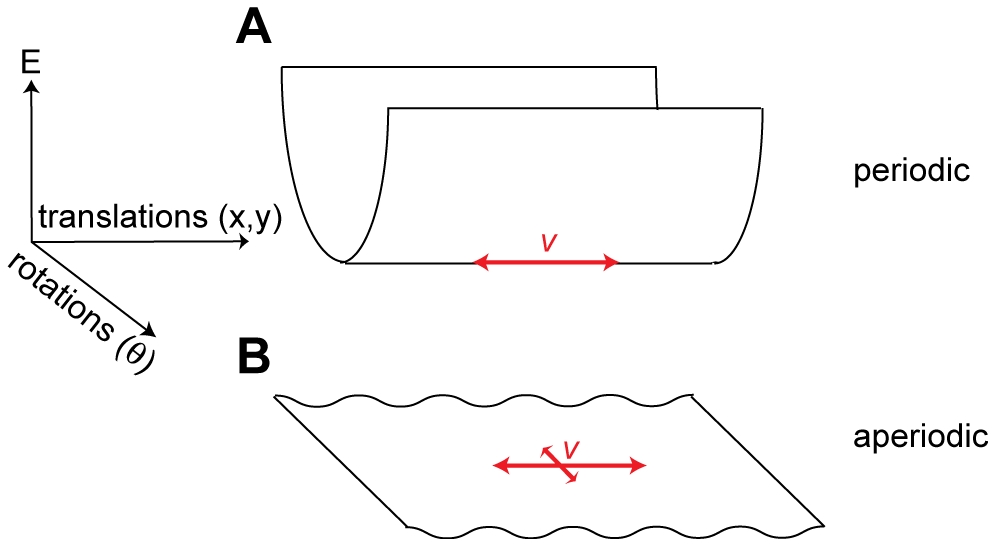}}
\caption{\textbf{The continuous attractor manifold.} {\bf A} Periodic network manifold: Points within the trough represent stable states of the network that will persist in the absence of perturbing inputs. If the network is placed at a state outside the trough, it will rapidly decay to a state within the trough. Points in the trough consist of continuous translations of the population-level pattern. Rotations, stretches, or other local or global deformations of the pattern lie outside the trough. Rat velocity inputs drive transitions between points in the trough (red arrow). {\bf B} Aperiodic network manifold: all rotations of a stable population pattern are energetically equivalent, and so form a continuous attractor manifold. Translations are not equivalent (rippled energy functional). Rat velocity inputs, when large enough to overcome the ripple, drive translations of the population pattern; however, the flat rotational mode means that the network can also rotate.}
\end{figure}


\begin{figure}[tbh]
\centerline{
\includegraphics[width=0.5 \textwidth]{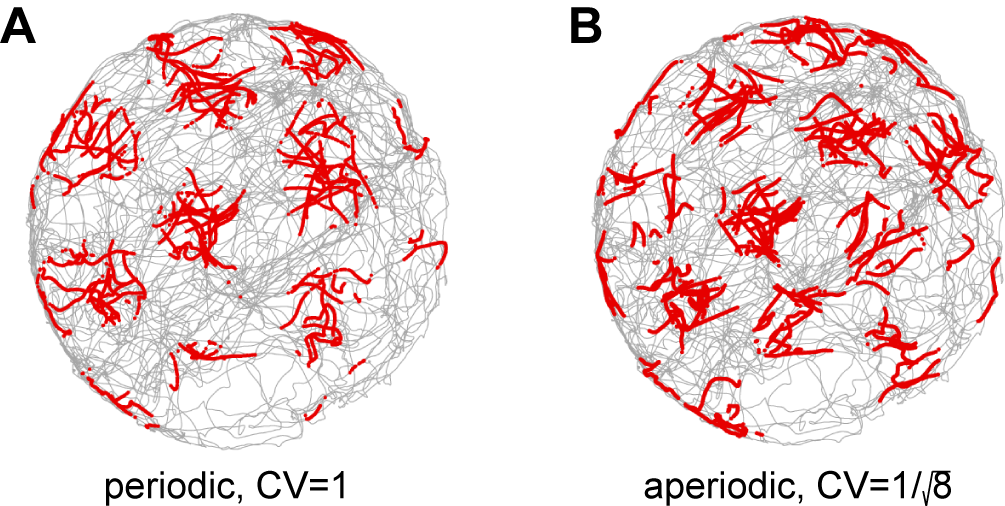}}

\caption{\textbf{Single neuron (SN) responses from stochastic spiking networks}. 
 {\bf A} SN response in a stochastic spiking periodic network. The parameters
and input velocity trajectory are as in Figure 2A--C, except that spiking 
is simulated explicitly and the spikes are generated by an inhomogeneous
Poisson process. {\bf B} SN response in a stochastic spiking aperiodic network.  The parameters are 
as in Figure 2D--F, except that spiking is simulated explicitly and the 
spikes are generated by a point process with a CV of $1/\sqrt{8}$ 
(see  \textit{Methods}). Each red dot represents a spike.}
\end{figure}


\begin{figure*}[bth]
\centerline{
\includegraphics[width=0.6 \textwidth]{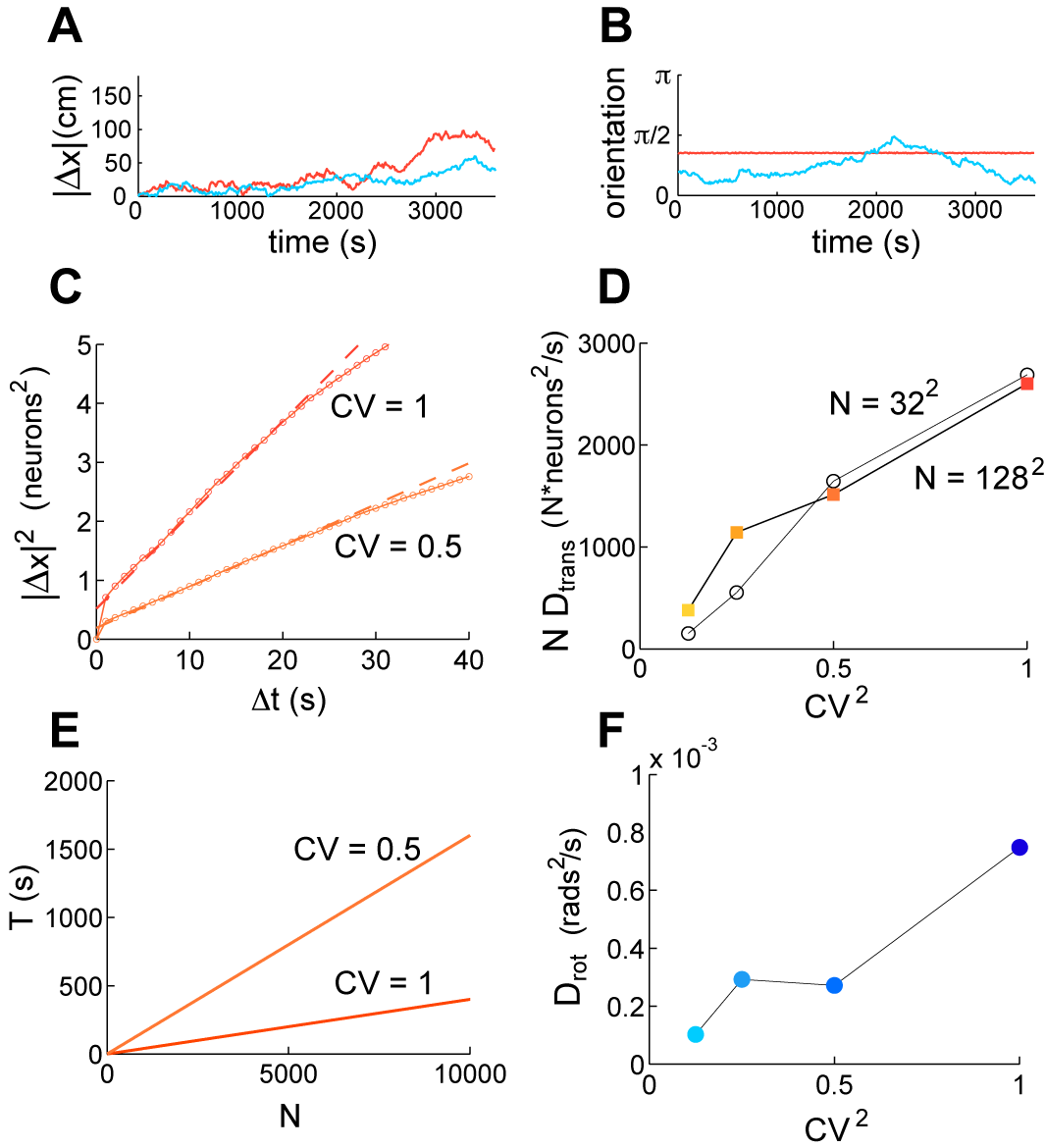}}
\caption{\textbf{Quantification of drift induced by neural stochasticity, in the absence of velocity inputs.} Orange (blue) curves are the results of simulations in (a)periodic networks. Successively darker shades (of orange or blue) represent simulations with successively higher neural variability (CV=$1/\sqrt{8}$, $1/\sqrt{4}$, $1/\sqrt{2}$, and $1$, respectively). Identical colors across panels represent simulations with identical network parameters. Velocity inputs are zero everywhere, and network size is $N=128^2$, except where stated otherwise. %
{\bf A} Phase  drift and {\bf B} angular drift of the periodic (orange, CV=1) and aperiodic (blue, CV=$1/\sqrt{8}$) networks. 
In {\bf A}, the drift in cm corresponds to a measured drift in neurons by assuming the same
gain factor as in the simulations with a trajectory, as in Fig. 5.
{\bf C} The summed square 2-d drift in position estimation as a function of elapsed time, for two different values of CV, in the absence of velocity inputs. The squared drift (small open circles) can be fit to straight lines (dashed) over 25 seconds (for longer times the traces deviate from the linear fit due to the finite time of the simulation), indicating that the process is diffusive. The slope of the line yields the diffusion constant $D_{\rm trans}$ for phase (translational) drift of the population pattern, in units of neurons\,$^2$/s. The same fitting procedure applied to the squared angular drift as a function of time yields the angular diffusion constant $D_{\rm rot}$.
{\bf D} Diffusion constants measured as in {\bf C}, for networks of varying size  and CV. The diffusion constant is approximately linear in CV$^2$, and in the number of neurons $N$. To demonstrate the linearity in $N$, the plots show $D$ multiplied by $N$, upon which the data for $N = 32^2$ and $N = 128^2$ approximately collapse onto a single curve.
{\bf E} An estimate of the time over which a periodic spiking network (with the same parameters as the corresponding points in C and D) can maintain a coherent grid cell response, plotted as a function of N, for two values of neural stochasticity. The estimate is based on taking the diffusion relationship $\Delta t = (\Delta x^2)/D_{\rm trans}$, and solving for the time when the average displacement $\Delta x$ is $10$ pixels, about half the population period, and estimating the diffusion constants from {\bf D} to be $ND \simeq 2500\, {\rm neurons}^2$/s. The coherence time scales like $\Delta t \propto \lambda^2 N/{\rm CV}^2$, where $\lambda$ is the period of the population pattern. 
{\bf F} Rotational diffusivity, $D_{\rm rot}$, in an aperiodic network of size 128 x 128 also increases linearly with CV$^2$. The diffusion constant was measured from  simulations lasting $20$ minutes.
}

\end{figure*}


\begin{figure*}[tbh]
\centerline{
\includegraphics[width=0.75 \textwidth]{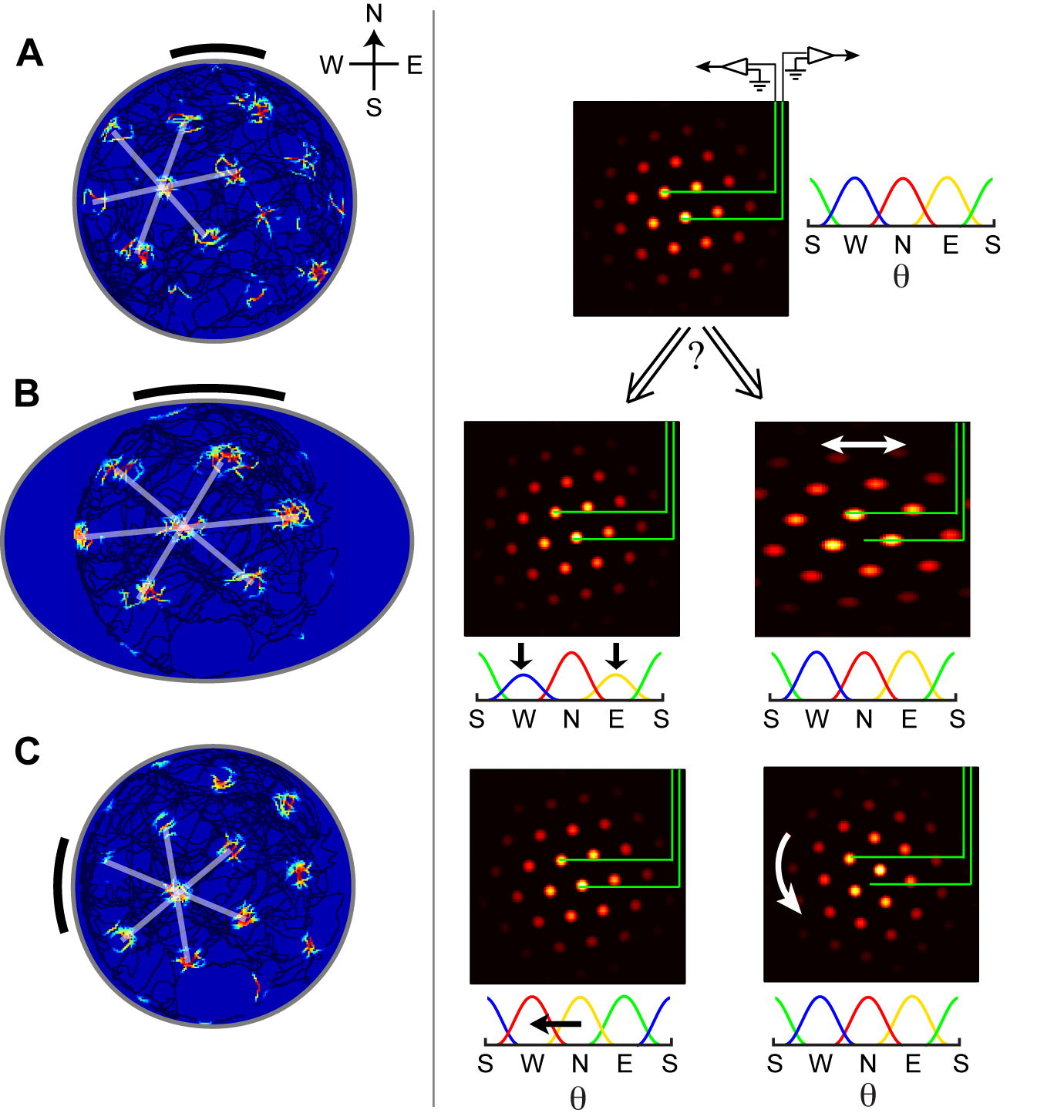}}
\caption{\textbf{Tests of the continuous attractor hypothesis.} Green lines represent the same fixed electrode locations in the neural population, across all plots. {\bf A} Left: Single-neuron response. Right: Input head direction/velocity tuning curves, and an instantaneous snapshot of the underlying population response, which together produced the SN response on the left. {\bf B} The SN grid (left) expands along one direction when the amplitude of the head direction/velocity inputs for that direction is lowered relative to other directions (right, first panel), while the population patterns remain unchanged. Alternatively, the same SN expansion could have been produced by keeping the amplitude of the head direction/velocity inputs fixed, if the population patterns were stretched (right, second panel). The latter scenario is inconsistent with the attractor hypothesis, because deformations of the pattern are not part of the attractor manifold. In the former (continuous attractor) scenario, the phase relationships between neurons is preserved despite the SN expansion; in the second, phase relationships must change. {\bf C} The SN grid (left) rotates if the head direction/velocity inputs to the network are rotated, while the population remains unchanged. The same rotation could have been produced by rotating the population pattern, but keeping the head direction/velocity inputs intact. The latter possibility is inconsistent with the attractor hypothesis. Again, the former (continuous attractor) scenario can be distinguished from the latter by whether phase relationships between neurons in the population are preserved. (SN plots and the left column of population responses were produced from a simulation with network parameters as in Figure 2\,D--F, and by appropriately scaling or rotating the velocity/head direction inputs. Right population plots are hypothetical.)}
\end{figure*}

\pagebreak


\renewcommand{\figurename}{Figure S}
\setcounter{figure}{0}

\begin{figure}[tbh]
\centerline{
\includegraphics[width=0.6 \textwidth]{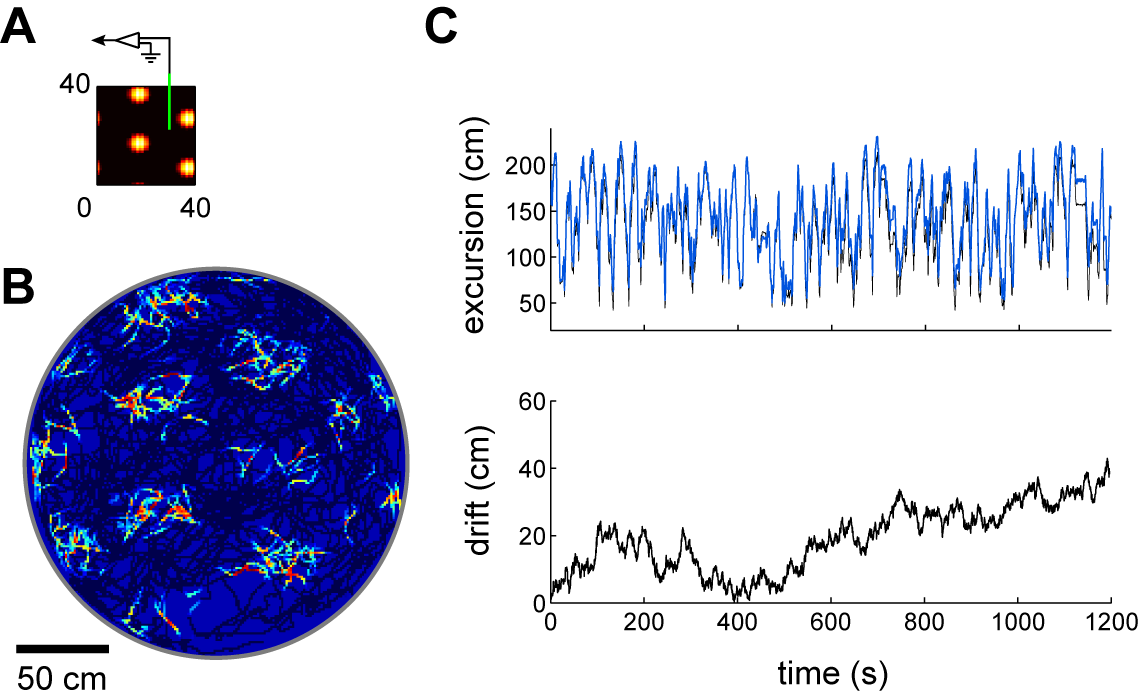}}
\caption{\textbf{Path integration and generation of grid cells in a small periodic network.} Simulation of network response, with velocity inputs corresponding to a rat's recorded trajectory in a 2m circular enclosure \citep{Haftingdatarelease}. The boundary conditions in the neural sheet are periodic as in Figure 2A--C, but the network size is smaller ($40^2$ network.) {\bf A} Instantaneous activity within the neural sheet (color represents the firing rate: black corresponds to vanishing rate). {\bf B} Grid cell response: average firing rate of a single neuron (located at the electrode tip in panel A), as a function of the rat's
position within the enclosure. {\bf C} Velocity integration in the network: Top: Actual distance of the rat from a fixed reference point (black), compared to the network's integrated position estimate, obtained by tracking the flow of the pattern in the population response (blue). The reference point is at the left-bottom corner of the square in which the circular enclosure is inscribed. Bottom: Accumulated difference between the integrated position estimate and the actual position.}
\end{figure}


\begin{figure}[tbh]
\centerline{
\includegraphics[width=0.6 \textwidth]{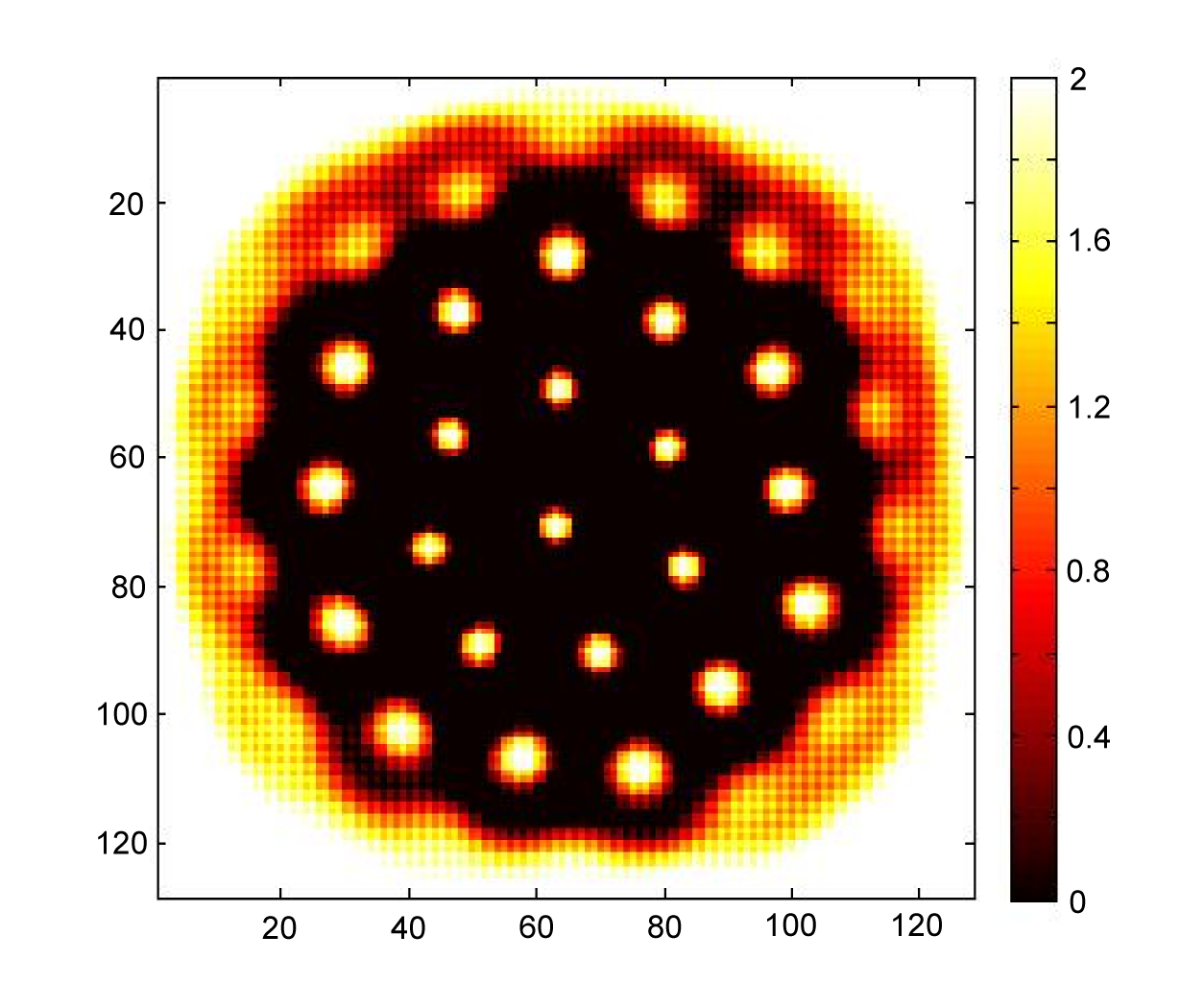}}
\caption{\textbf{Population pattern in an aperiodic network with a modulation of weights.} The steady-state pattern in a network where the strengths of the outgoing weights from each neuron are modulated based on the neuron's location in the sheet, according to the envelope function of Equation 5. 
The external input is spatially uniform. All parameters are identical to the simulation of Figure 2 D, except that the modulation envelope is applied to the weights
instead of to the inputs. The formed pattern is distorted at the edges, with neurons along the edge tending to be uniformly active.}
\end{figure}


\begin{figure}[tbh]
\centerline{
\includegraphics[width=0.75 \textwidth]{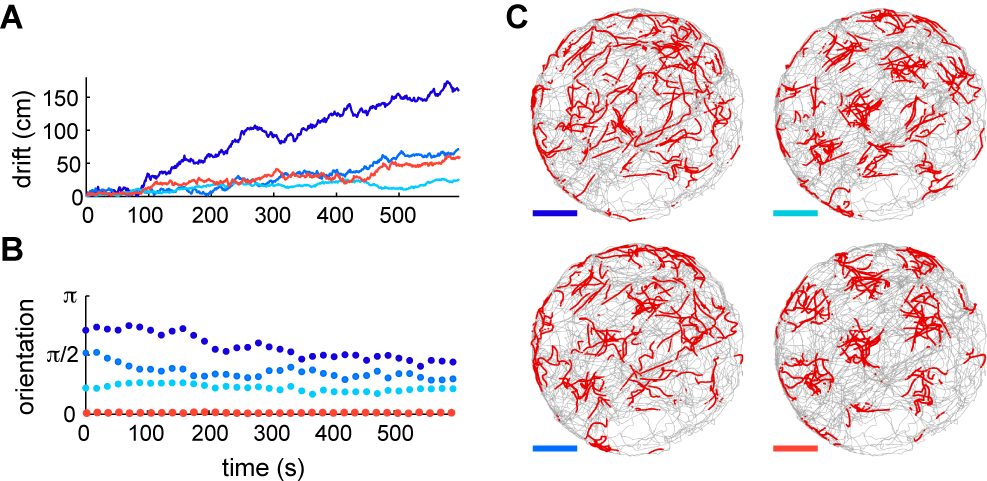}}
\caption{\textbf{Path integration in periodic and aperiodic stochastic spiking networks.} 
Simulation of network response, with velocity inputs corresponding to a rat's recorded trajectory in a 2m circular enclosure \citep{Haftingdatarelease}, in stochastic spiking networks. Results are shown for a periodic network with CV = 1 (orange), and for aperiodic networks, where successively darker shades of blue represent simulations with successively higher neural CV (CV=$1/\sqrt{8}$, $1/\sqrt{4}$, and $1$, respectively). All other parameters are as in Figure~5. Colors represent the same network parameters as in Figure~6, which describes drift in the absence of velocity inputs.
{\bf A} Accumulated difference between the integrated position estimate and the rat's actual position. {\bf B} Orientation of the network pattern as a function of time. {\bf C} Responses of a single neuron over a rat's recorded trajectory, over $10$ minutes. Each red dot represents a spike. Color of bars represent the same simulation parameters as in ${\bf A}$ and ${\bf B}$. Top-left, Aperiodic network with 
CV$=1$,  Bottom-left, CV$=1/\sqrt{4}$, Top-right, 
CV$=1/\sqrt{8}$ (reproduced from Fig.~5), Bottom right, aperiodic network with CV=$1$ (reproduced from Figure~5).
}
\end{figure}


\begin{figure}[tbh]
\scalebox{0.75}{
\includegraphics{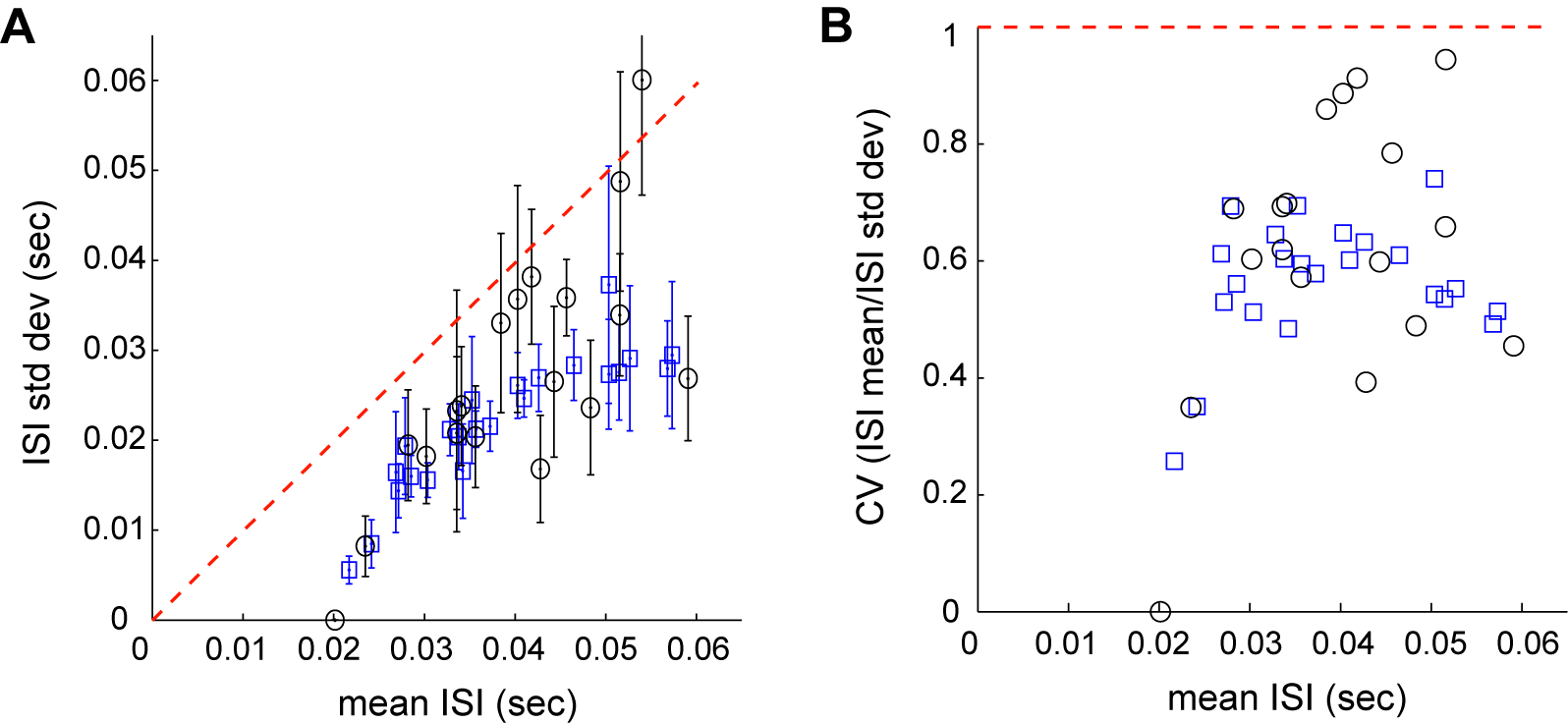}}
\caption{\textbf{Stochasticity of recorded dMEC neurons.}
\small
\textbf{A} Standard deviation ($\sigma$) of the inter-spike interval (ISI) distribution plotted against the mean ISI, for various values of the mean ISI. Data points from multiple simultaneously recorded cells (from a single electrode) in dMEC \citep{Haftingdatarelease} are pooled to produce this plot. Black circles, method (1). Blue squares, method (2) (see below). The red dashed line corresponds to statistics that would be obtained from a homogeneous Poisson process at each mean ISI value.
\textbf{B} The coefficient of variation (CV=$\sigma$(ISI)/$\mu$(ISI)) plotted as a function of spiking frequency. The red dashed line corresponds to the CV of a Poisson process.
{\bf Estimation of CV in neural data}: The CV is a normalized measure of the variation in the inter-spike intervals in a spike train firing at a constant rate. To estimate the CV, we thus have to identify intervals of relatively constant firing rate. This is made complicated by the fact that in the stimulus and behavioral conditions prevailing during the recordings (the rat is randomly running around the enclosure foraging for randomly scattered food while landmarks move into or out of view), there are no designated regions of stimulus or response constancy. We used two methods to identify regions of constant mean firing rate: (1) Identify blocks of low-velocity intervals where $|{\bf v}|<v_{\rm cutoff}=8$cm/s, which are of duration larger than $T_{\rm v} = 4$s. We found no blocks where the integrated displacement was more than $\lambda/4$cm, meaning that the intervals represented traverses of approximately one blob diameter or less, with the typical distance being much shorter. Thus, the rat is likely to be either on or off a blob for the entire duration of a block, and should have a roughly constant underlying firing rate. (2) Identify high-rate blocks where the rate is higher than some upper cutoff threshold (to locate on-blob episodes), with $r_{\rm ISI}(t)>r_{\rm high}$ for each time in the block. Only those high-rate blocks of duration longer than $T_{\rm r}$ were retained. $r_{\rm ISI}$ is the instantaneous firing rate, computed as the reciprocal of the inter-spike interval of adjacent spikes. $r_{\rm high}=10$Hz was chosen to be large enough to exclude all intervals except those where the rat is clearly on a blob for the recorded cell. In all the above, the minimum interval duration $T_{\rm r}=5$s was chosen to eliminate random (non)spike events that momentarily change the rate without reflecting an actual change in the underlying firing rate of the cell, while capturing as many intervals as possible for ISI analysis. In each of methods (1) or (2), we compute $\mu$(ISI) and $\sigma$(ISI) for each block as a single data-point. Next, we bin together data points with the same rate (in bins of 1 Hz), pooling across all cells (this is reasonable because each cell individually has very similar statistics as the collection). The two methods (1) and (2) are complementary in the sense that interval sampling is based in the first case on rat velocity, and in the second case by rate-based on-blob or off-blob considerations. Neither method guarantees that the underlying firing rate within one interval is constant. However, the two methods yield consistent results, and thus add a measure of confidence to the analysis.
}
\end{figure}


\begin{figure}[tbh]
\centerline{
\includegraphics[width=0.9 \textwidth]{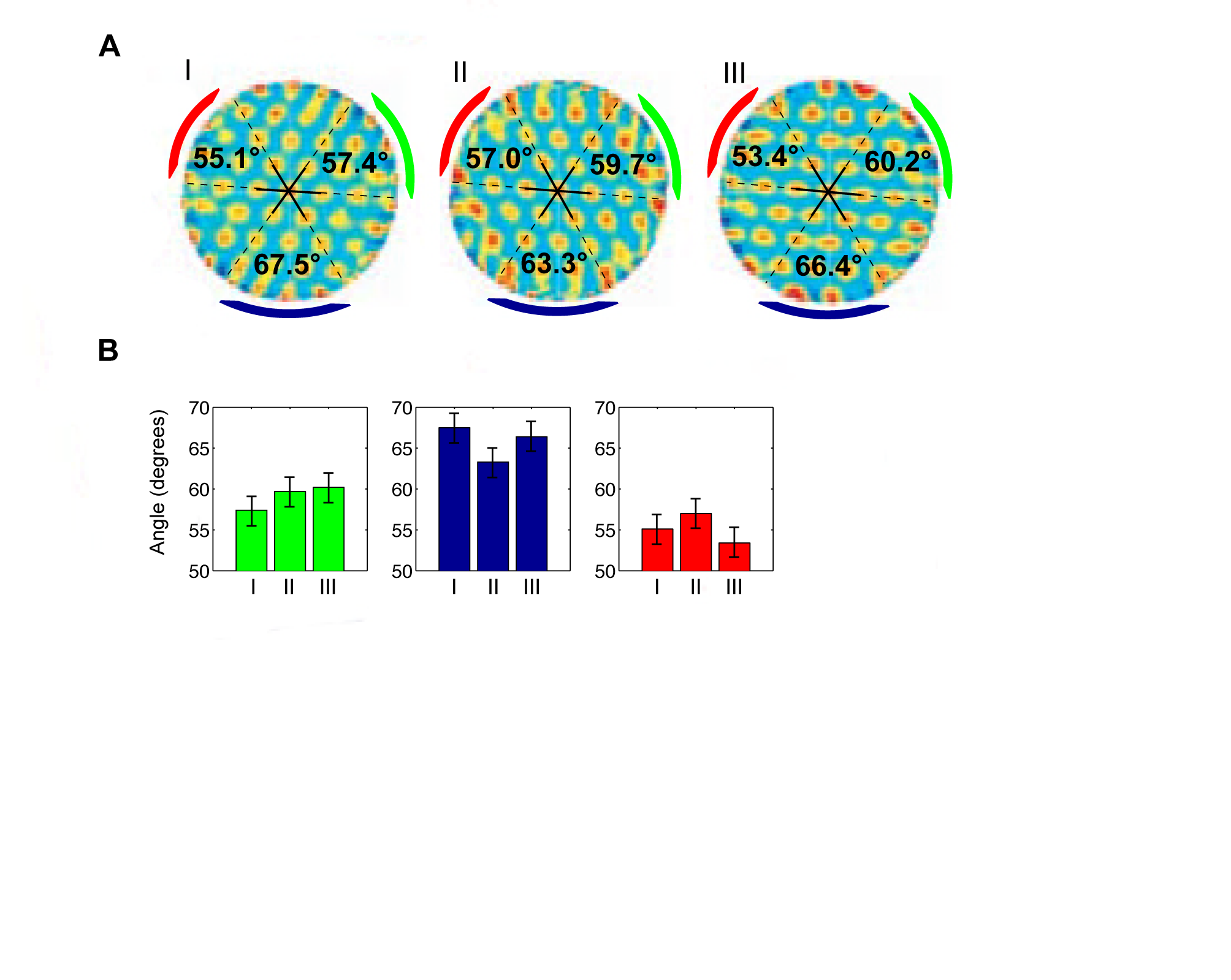}}
\caption{\textbf{Deviations from a perfect triangular lattice in existing measurements.} {\bf A} Comparison of grid correlation functions from three simultaneously recorded cells, adapted from \citep{Hafting05}. The black lines were passed between pairs of peaks in the correlation function. Each pair consists of two opposing peaks, from the six closest peaks to the origin. Measured angles between the lattice vectors, shown in the plot and in the bar plot, {\bf B}, show a consistent bias from 60¡ in the three cells. We estimate the measurement error at about $\pm 2^{o}$. The measured lengths of the black segments, in arbitrary pixel units, are: 28.8, 27.2, 25.1 (I); 28.8, 27.5, 25.9 (II); 29.2, 28.7, 26.1 (III), with an estimated measurement error of +-1. This example is limited by the low resolution images adapted from \citep{Hafting05} and is meant primarily as a demonstration of possible deviations from a perfect triangular lattice, and how they can be measured. We believe that the question of whether such deviations occur consistently in cells sharing the same grid period calls for a more systematic study.}
\end{figure}

\end{document}